\documentclass[
reprint,
 amsmath,amssymb,
 aps,
]{revtex4-1}

\usepackage[english]{babel}
\usepackage[T1]{fontenc}
\usepackage[utf8]{inputenc}
\usepackage{bm}
\usepackage{mathtools}
\usepackage{color}
\usepackage{enumerate}
\usepackage[hidelinks]{hyperref}

\newcommand{\D}{\text d}
\newcommand{\ImI}{\text i}
\newcommand{\EuE}{\text e}

\newcommand{\intBZpi}[1]{\int\displaylimits_{\mathclap{\text{BZ}}}\frac{\D^3 #1}{(2\pi)^3}}
\newcommand{\intBZpiT}[3]{\int\displaylimits_{\mathclap{(\text{BZ})^3}}\frac{\D^3 #1}{(2\pi)^3} \frac{\D^3 #2}{(2\pi)^3} \frac{\D^3 #3}{(2\pi)^3}}
\newcommand{\sumBZm}[3]{\sum_{\bm #1_#2 ... \bm #1_#3}^{\text{BZ}}}

\definecolor{darkred}{rgb}{0.4, 0.0, 0.0}
\newcommand{\revcolI}[1]{{#1}}

\hypersetup{
  colorlinks   = true, %Colours links instead of ugly boxes
  urlcolor     = blue, %Colour for external hyperlinks
  linkcolor    = blue, %Colour of internal links
  citecolor   = red    %Colour of citations
}

\begin{document}

%\title{Quartic scaling second-order M\o eller-Plesset Perturbation Theory\\for periodic systems: A highly parallelized algorithm in the plane wave basis}
\title{Quartic scaling MP2 for solids:\\A highly parallelized algorithm in the plane wave basis}
\author{Tobias Sch\"afer}
\author{Benjamin Ramberger}
\author{Georg Kresse}
\affiliation{University of Vienna, Faculty of Physics and Center for Computational Materials Science, Sensengasse 8/12, A-1090 Vienna, Austria}

%\date{\today}

\begin{abstract}
We present a low-complexity algorithm to calculate the correlation energy of periodic systems in second-order M\o ller-Plesset perturbation theory (MP2). In contrast to previous approximation-free MP2 codes, our implementation possesses a quartic scaling, $\mathcal O (N^4)$, with respect to the system size $N$ and offers an almost ideal parallelization efficiency. The general issue that the correlation energy converges slowly with the number of basis functions is \revcolI{eased} by an internal basis set extrapolation. \revcolI{The key concept to reduce the scaling is to eliminate all summations over virtual orbitals which can be elegantly achieved in the Laplace transformed MP2 (LTMP2) formulation using plane wave basis sets and Fast Fourier transforms.} 
%Compared to the previous, quintic scaling, MP2 code in \textsc{vasp} our lower scaling code has a slightly higher prefactor, but is favourable for system containing more than 10 to 30 atoms. 
%We show that the MP2 energy of a LiH supercell containing 128 atoms can readily be calculated with modest computer power and we estimate that X atoms with Y basis functions can be calculated in Z using 100 000 CPU cores.
Analogously, this approach could allow to calculate second order screened exchange (SOSEX) as well as particle-hole ladder diagrams with a similar low complexity. Hence, the presented method can be considered as a step towards systematically improved correlation energies.
\end{abstract}

\maketitle

% Vienna Ab initio Simulation Package (VASP) \cite{Kresse1993,VASP2}
% reciprocal lattice vector (plane wave)
% The informed reader should skip the theory part and only read the schematic illustration

% TODO:
% change G <--> G'
% make decomposition better visible

%%%%%%%%%%%%%%%%%%%%%%%%%%%%%%%%%%%%%%%%%%%%%%%%%%%%%%%%%%%%%%%%%
\section{Introduction}
%%%%%%%%%%%%%%%%%%%%%%%%%%%%%%%%%%%%%%%%%%%%%%%%%%%%%%%%%%%%%%%%%

%Why MP2? History. State of the Art. Our approach. Strucutre of the paper. Blabla.

When calculating the ground state energy of matter in a perturbative approach, the second-order M\o ller-Plesset perturbation theory (MP2) \cite{Moeller1934} is the lowest order correction to the well-established Hartree-Fock (HF) approximation. This correction includes electron correlation and non-covalent effects like van der Waals interaction, making MP2 very attractive for ab initio calculations of molecules and nonmetallic solids. However, in the traditional MP2 formulation \cite{Szabo1996}, the improvements compared to HF come along with a very high price of computational effort. The scaling of the computation time with respect to the system size is quintic, $\mathcal O(N^5)$, if no further approximation is made.

To overcome this steep scaling, several attempts have been made for finite systems like molecules \cite{Ayala1999, Schutz1999, Saebo2001, Doser2009, Nagy2016}, where linear scaling codes, $\mathcal O(N)$, could be achieved using approximations based on the locality of one-electron orbitals, local MP2 (LMP2), and Gaussian basis sets. 
%\revcolI{Nonetheless, it is still difficult to achieve high accuracy for large three dimensional systems, i.e. to achieve convergence with respect to the basis sets size.}
\revcolI{Notwithstanding the linear scaling, it remains computationally demanding to achieve high accuracy and converged basis sets for three dimensional systems with small band gaps.}
%Nonetheless, it is generally agreed that these methods can suffer from accuracy problems \cite{Subotnik2008} and slow down considerably when improved diffuse basis sets are used.

For periodic systems like crystalline solids, growing interest in MP2 calculations can be observed \cite{Manby2006,Casassa2008,Halo2009,Halo2009a,Erba2009,Erba2011,Casassa2012,Fabiano2009,Maschio2010,Maschio2010a,Schwerdtfeger2010,Nanda2012,Stodt2012,Goltl2012,Muller2013,DelBen2013,DelBen2014,DelBen2015,Torabi2014,Hammerschmidt2015,Kaawar2016}. Like for molecules plenty of MP2 implementations are available for periodic systems nowadays. While applications for specific extended 1D and 2D systems go back to Suhai \cite{Suhai1983}, and Sun and Bartlett \cite{Sun1996}, the first general purpose MP2 computer program for periodic systems was \textsc{cryscor} by \revcolI{the Pisani and Sch\"utz group} \cite{Pisani2005,Pisani2008}. Based on the LMP2 approach for molecules, \textsc{cryscor} inherits the linear scaling, $\mathcal O(N)$, but also the mentioned accuracy and basis set issues\revcolI{, although significant progress has been made in recent years \cite{Usvyat2015}.} \revcolI{The first plane wave based MP2 implementation for periodic systems was made available in \textsc{vasp} \cite{Kresse1993,Kresse1999} by Marsman, Grüneis et al. \cite{Marsman2009, Gruneis2010}.} Yet, this implementation sustains the unfavorable quintic scaling, $\mathcal O(N^5)$, making it difficult to treat large systems with more than 64 atoms, both regarding memory and computation time.\\
Further methods for periodic systems balancing computational effort and accuracy rely on the resolution-of-identity approximation (RI) \cite{Katouda2010}, both RI and localized atomic orbitals \cite{Izmaylov2008,Maschio2007,Usvyat2006}, or real-space Monte Carlo integration of a Green's function based MP2 formulation~\cite{Willow2012, Willow2014}.\\
Finally, a high performance code for finite and periodic MP2 calculations became available quite recently, providing a high parallelization efficiency. Implemented in \textsc{cp2k} by VandeVondele and co-workers \cite{DelBen2012}, this high performance approach possesses a reduced prefactor for a, still, largely quintic scaling, $\mathcal O(N^5)$.

Here we present a novel implementation of a high performance algorithm for exact MP2 calculations of periodic systems that provides a very high parallelization efficiency with low memory requirements and the computation time scales only quartic with the system size, $\mathcal O(N^4)$. The lower scaling is achieved by the Laplace-transformed MP2 \cite{Almlof1991} formulation (LTMP2) \revcolI{and Fast Fourier transformations}, allowing for a presummation over all virtual orbitals. The method is implemented in \textsc{vasp}. The high parallelization efficiency is attained by dividing the set of all plane waves over all CPUs, leading to a communication-free distribution of an outer plane wave loop.

In this paper we begin with a theoretical part, Sec. \ref{sec:Theory}, containing a brief introduction to the canonical MP2 formulation, a schematic summary of the main strategy to reduce the scaling, and a comprehensive derivation of the quartic scaling LTMP2 formulation for periodic systems \revcolI{including k-point sampling}. In Sec. \ref{sec:Impl} the implementation is described regarding the parallelization strategy and the internal basis set extrapolation. We also provide pseudocode for the serial and parallel version. Benchmark calculations can be found in Sec. \ref{sec:Bench}. We show the measured scaling behavior in both computation time and memory, the parallelization efficiency, as well as the numerical agreement of the resulting MP2 energies of our new code and the previous MP2 code of \textsc{vasp}. Lithium hydride (LiH) and methane (CH${}_4$) in a chabazite cage serve as benchmark systems.

%We report that supercells with up to 128 atoms with 5600 basis functions (crystal LiH) can be calculated n less than 19 hours with modest computer power of only 64 cores.
%or within 26 minutes using 2048 cores.

% 432 Atoms (3x3x3 NH3 supercell) or 20 000 basis functions feasible on 100 000 cores in less than two hours.

% However all these approaches relie on special assuptioms of the symmetry of the system and, thus, are not general and a compromise between accuracy and computation time.

%For finit systems like molecules 

%Here we present an algorithm that has a lower scaling, $\mathcal O(N^4)$, and a extremley high parallelization efficiency.

%\section*{ToDo}
%\begin{itemize}
%\item intro (referring to vertex corrections?)
%\item interchange $\bm G$ and $\bm G'$ in the periodic MP2 part
%\item nur irreducible BZ?
%\item more references
%\item assign a name to the transformed states $|w\rangle$ 
%\end{itemize}

%%%%%%%%%%%%%%%%%%%%%%%%%%%%%%%%%%%%%%%%%%%%%%%%%%%%%%%%%%%%%%%%%
\section{Theory} \label{sec:Theory}
%%%%%%%%%%%%%%%%%%%%%%%%%%%%%%%%%%%%%%%%%%%%%%%%%%%%%%%%%%%%%%%%%

In M\o ller-Plesset perturbation theory \cite{Moeller1934} the correlation energy $E_c$ is estimated by means of a Rayleigh-Schr\"odinger perturbation series of the ground state energy 
\begin{equation}
E=E^{(0)}+E^{(1)}+E^{(2)}+... \;.
\end{equation}
The correlation energy is conventionally defined as the the difference between the ground state energy and the Hartree-Fock energy, since the latter neglects correlation effects: $E_c = E - E_{\text{HF}}$. Starting from a Hartree-Fock Hamiltonian, $H_\text{HF}$, the full Hamiltonian, $H$, is considered as the Hartree-Fock Hamiltonian complemented by a perturbation: 
\begin{equation}
H = H_\text{HF} + H_c \;,\quad  H_c = H - H_{\text{HF}} \;.
\end{equation}
The perturbation $H_c$ contains the full electron-electron interaction minus the electron-mean field interaction. Applying the common formulae of Rayleigh-Schr\"odinger perturbation theory, the sum of the zeroth and first order term of the perturbation series turns out to be the Hartree-Fock energy: $E_{\text{HF}}=E^{(0)}+E^{(1)}$. Thus the leading contribution to the correlation energy is given by the second order term (MP2 energy): $E_\text c = E^{(2)}+...$ . The textbook equation \cite{Szabo1996} reads:
\begin{equation}
E^{(2)} = \frac 1 2 \sum_{ij}^{\text{occ.}} \sum_{ab}^{\text{virt.}} \frac{\langle ij | ab \rangle[\langle ab| ij\rangle - \langle ab| ji\rangle] }{\varepsilon_{i} + \varepsilon_{j} - \varepsilon_{a} - \varepsilon_{ b } } \;. \label{eq:CanMP2}
\end{equation}
The orbitals $|i\rangle$ / $|a\rangle$ and energies $\varepsilon_i$ / $\varepsilon_a$ are the occupied / virtual solutions of the Hartree-Fock equations respectively. With $\langle ab| ij\rangle$ we identify a two-electron integral,
\begin{equation}
\langle ij | ab \rangle = \int \D^3 r \int \D^3 r' \; \frac{\varphi_i^*(\bm r) \varphi_j^*(\bm r')\varphi_a(\bm r)\varphi_b(\bm r') }{|\bm r - \bm r'|}  \;,
\end{equation}
which is simply a matrix element of the two-electron Coulomb operator. The first part in (\ref{eq:CanMP2}), containing $\langle ij|ab\rangle\langle ab | ij \rangle$, is often called direct MP2 energy $E^{(2)}_\text{d}$ and the second part, containing$-\langle ij|ab\rangle\langle ab | ji \rangle$, is often called exchange MP2 energy $E^{(2)}_\text{x}$.
Hartree atomic units are used throughout the paper.

\subsection{Reducing the computational cost} \label{sec:brief_strategy}

In order to give a clear overview of the main strategy to reduce the computational cost of the MP2 energy calculation, we limit ourselves to a schematic description (neglecting  spin and periodicity and considering only  $\bm\Gamma$-point sampling of the Brillouin zone) in this section. A proper and comprehensive derivation is given in \ref{subsec:periodMP2}. Furthermore, we restrict to the computationally most expensive part, the exchange term of the MP2 energy (\ref{eq:CanMP2}), given by
\begin{equation}
E^{(2)}_\text{x} = -\frac 1 2 \sum_{ij}^{\text{occ.}} \sum_{ab}^{\text{virt.}} \frac{\langle ij | ab \rangle \langle ab| ji\rangle }{\varepsilon_{i} + \varepsilon_{j} - \varepsilon_{a} - \varepsilon_{ b } } \;. \label{eq:SCHMP2}
\end{equation}
In this canonical form the (direct/exchange) MP2 energy has a quintic scaling with the system size. \revcolI{This is due to the fact that the two-electron integrals $\langle ij | ab \rangle$ and $\langle ab | ji \rangle$ have to be evaluated for all combinations of $i,j,a,b$, where the computation of $\langle ij | ab \rangle$ and $\langle ab | ji \rangle  $, for given $i,j,a,b$, scales linear with the system size for plane wave codes. This linear scaling can be seen when the Coulomb operator in the two-electron integrals is written in reciprocal space: 
\begin{equation}
\langle ij | ab \rangle = \frac 1 \Omega \sum_{\bm G} \frac{4\pi}{\bm G^2}\,  \langle i | \EuE^{-\ImI\bm G \hat {\bm r}} | a \rangle \langle j | \EuE^{\ImI\bm G \hat {\bm r}} | b \rangle  \;, \label{eq:SCH2EI} 
\end{equation}
where $\Omega$ is the volume of the system, and the overlap densities  are a Fourier transform $\mathcal F$ of HF orbitals,
\begin{equation}
\langle i | \EuE^{-\ImI\bm G \hat {\bm r}} | a \rangle = \int \D^3 r \; \varphi^*_i(\bm r) \varphi_a(\bm r) \EuE^{-\ImI \bm G \bm r} = \mathcal{F}[\varphi_i^*\varphi_a](\bm G) \;. \label{eq:SCHovrlp}
\end{equation}
The orbitals,
\begin{equation}
|i\rangle = \sum_{\bm G} \widetilde \varphi_i(\bm G) \,|\bm G\rangle\;,
\end{equation}
are represented in terms of the plane wave coefficients $\widetilde \varphi_i(\bm G)$, where $\langle \bm r | \bm G \rangle = \Omega^{-\frac 1 2} \,\EuE^{\ImI\bm G\bm r}$ is a plane wave. In practice, the basis for the orbitals is
truncated at a plane wave cutoff $E_{\text{cut}}$ ({\tt ENCUT} in \textsc{vasp}) and only plane waves observing 
\begin{equation}
 \frac {\bm G^2} 2 \le E_{\text{cut}}
\end{equation}
are used. For the overlap densities in Eq. (\ref{eq:SCHovrlp}) a different cutoff $E^{\text{aux}}_{\text{cut}}$ ({\tt ENCUTGW} in \textsc{vasp}) is used which is commonly chosen as $2/3 E_{\text{cut}}$ in \textsc{vasp}. This second basis
set is analogous to the auxiliary basis sets used in Gaussian type orbital codes. As will be discussed
later, the internal basis set extrapolation is performed with respect to the auxiliary basis set size. 
This approach has been successfully used in the random phase approximation (RPA) in the past \cite{Harl2008}.
The orbitals in real space, $\varphi_i(\bm r)$, are easily obtained by switching from $\bm G$ to $\bm r$ by an inverse Fourier transform, $\mathcal F^{-1}[\widetilde \varphi_i](\bm r)$.
%\begin{equation}
%\varphi_i(\bm r) = (2\pi)^3 \mathcal F^{-1}[\widetilde \varphi_i](\bm r)\;.
%\end{equation}
In plane wave codes the Fourier transform is replaced by a Fast Fourier transform (\texttt{FFT}). If the overlap densities 
$\langle i | \EuE^{-\ImI\bm G \hat {\bm r}} | a \rangle$
are precalculated  and stored, the computation time of (\ref{eq:SCHMP2}) scales as $\mathcal O(N_i^2N_a^2N_G)$, where $N_i$ is the number of occupied orbitals and $N_a$ is the number of virtual orbitals,
and $N_G$ is the number of basis functions in the auxiliary basis set. 
This is the canonical quintic scaling in a plane wave basis.}

Our strategy to reduce the computational cost consists of the idea to decouple the sums over the bands $i,j,a,b$ such that the summation over all virtual bands $a,b$ can be performed in advance. The decoupling of the band summation is achieved by a Laplace transformation of the energy denominator \cite{Almlof1991}: 
\begin{equation}
\frac{1}{\varepsilon_    {i} + \varepsilon_{j} - \varepsilon_{a} - \varepsilon_{ b }} = -\int_0^\infty \D \tau \; \EuE^{(\varepsilon_    {i} + \varepsilon_{j} - \varepsilon_{a} - \varepsilon_{ b })\tau} \;. \label{eq:SCHLT}
\end{equation}
It is then possible to rewrite the summand in Eq. (\ref{eq:SCHMP2}) such that the sum over $a$ and $b$ can be performed first:
\begin{align}
&-\sum_{ab}^{\text{virt.}}\frac{\langle ij | ab \rangle \langle ab| ji\rangle }{\varepsilon_{i} + \varepsilon_{j} - \varepsilon_{a} - \varepsilon_{ b } } \nonumber\\[10pt]
=&\;\frac{1}{\Omega^2} \sum_{ab}^{\text{virt.}}  \int_0^\infty \D \tau \; \EuE^{(\varepsilon_    {i} + \varepsilon_{j} - \varepsilon_{a} - \varepsilon_{ b })\tau}\sum_{\bm G \bm G'} \frac{4\pi}{\bm G^2} \frac{4\pi}{\bm G'^2}  \nonumber  \\
&\times\langle i | \EuE^{-\ImI\bm G' \hat {\bm r}} | a \rangle \langle j | \EuE ^{\ImI\bm G' \hat {\bm r}} | b \rangle  \langle a | \EuE^{-\ImI\bm G \hat {\bm r}} | j \rangle \langle b | \EuE^{\ImI\bm G \hat {\bm r}} | i \rangle \nonumber\\[10pt]
=&\;\frac{1}{\Omega^2} \int_0^\infty \D \tau \; \sum_{\bm G \bm G'} \frac{4\pi}{\bm G^2} \frac{4\pi}{\bm G'^2} \nonumber \\
&\times \langle i | \EuE^{-\ImI\bm G' \hat {\bm r}}  \sum_{a}^{\text{virt.}}\left[   \langle a | \EuE^{-\ImI\bm G \hat {\bm r}} | j \rangle \EuE^{(\varepsilon_{j} - \varepsilon_{a})\tau} | a \rangle \right] \nonumber \\
&\times \langle j | \EuE^{+\ImI\bm G' \hat {\bm r}}  \sum_{b}^{\text{virt.}}\left[   \langle b | \EuE^{+\ImI\bm G \hat {\bm r}} | i \rangle \,\EuE^{(\varepsilon_{i} - \varepsilon_{b})\tau} \,| b \rangle \right]  \;.
\end{align}
The squared brackets indicate how the sum over $a$ and $b$ can be performed in advance for each considered $\bm G$ and $\tau$. This summation defines transformed states:
\begin{equation}
\Big | w_j^{(\bm G \tau)} \Big \rangle  = \sum_a C_{ja}^{(\bm G \tau)} \, | a \rangle \;,  \label{eq:SCHnewstate}  \\
\end{equation}
where the transformation matrix, for given $\bm G$ and $\tau$, reads:
\begin{equation}
C_{ja}^{(\bm G \tau)}  = \EuE^{(\varepsilon_j-\varepsilon_a)\tau} \langle a | \EuE^{-\ImI\bm G \hat {\bm r}} | j \rangle \label{eq:SCHtransmatrix}  \;.
\end{equation}
Then the MP2 exchange energy takes the form of a Fock-like energy:
\begin{equation}
E^{(2)}_\text{x} = \frac 1 2 \frac{1}{\Omega^2} \int_0^\infty \D \tau  \sum_{\bm G} \frac{4\pi}{\bm G^2} \sum_{ij}^{\text{occ.}}  \Big\langle ij \Big|  w_j^{(\bm G \tau)}  w_i^{(-\bm G, \tau)} \Big\rangle \;. \label{eq:SCHfocklike}
\end{equation}
\revcolI{Here the two-electron integral $\langle ij |  w_j w_i\rangle$ can again be calculated with linear scaling (see Eq. \ref{eq:SCH2EI}) for given $i,j,\bm G,\tau$:
\begin{multline}
\Big\langle ij \Big|  w_j^{(\bm G \tau)}  w_i^{(-\bm G, \tau)} \Big\rangle\\
= \frac 1 \Omega \sum_{\bm G'} \frac{4\pi}{\bm G'^2}\,  \big\langle i \big| \EuE^{-\ImI\bm G' \hat {\bm r}} \big| w_j^{(\bm G \tau)} \big\rangle \big\langle j \big| \EuE^{\ImI\bm G' \hat {\bm r}} \big| w_i^{(-\bm G, \tau)} \big\rangle\;.
\end{multline}
Since the $\tau$-integration is performed by a quadrature where the number of $\tau$-points is largely independent of the system size, the scaling of the exchange term, $E^{(2)}_\text{x}$, is reduced to $\mathcal O (N_i^2N_{\bm G}N_{\texttt{FFT}}\ln N_{\texttt{FFT}})$, where $N_{\bm G}$ is the number of plane waves in the auxiliary basis and $N_{\texttt{FFT}}$ is the number of \texttt{FFT} grid points. Hence the presummation over the virtual bands saves costs of $\mathcal O(N_a^2)$ at the expense of $\mathcal O(N_{\bm G})$ leading to an improvement of one order of magnitude.  For the direct term, $E^{(2)}_\text{d}$, the scaling is reduced to $\mathcal O (N_iN_{\bm G}N_{\texttt{FFT}}\ln N_{\texttt{FFT}})$  using the same technique. However, the calculation of the transformed states (\ref{eq:SCHnewstate}) also possesses a quartic scaling such that the overall scaling of the algorithm is quartic for both the direct and the exchange MP2 energy.}

\subsection{MP2 for periodic systems}\label{subsec:periodMP2}

In the following subsections we elaborate the aforesaid strategy to reduce the scaling of the MP2 energy for a periodic system in detail. For a periodic system the MP2 energy per unit cell is given \cite{Suhai1983} by 
\begin{equation}
E^{(2)} = \frac 1 2 \frac 1 N  \sum_{IJ}^{\text{occ.}} \sum_{AB}^{\text{virt.}} \frac{\langle I J | A B\rangle \big[\langle A B | I J\rangle - \langle A B | J I\rangle\big] }{\varepsilon_I + \varepsilon_J - \varepsilon_A - \varepsilon_B}\;.
\end{equation}
Here $N$ is the number of unit cells composing the entire system. The capital letters $I,J,A,B$ are composite indices representing all quantum numbers. Still $I$ and $J$ run over all occupied Hartree-Fock one-electron states whereas $A$ and $B$ pass through all virtual (unoccupied) states of the system. The $\varepsilon_I$'s and $\varepsilon_A$'s are the occupied and virtual Hartree-Fock one-electron energies respectively. 

We assume the system to be periodic (e.g. a periodic 3D lattice). Consequently the quantum numbers are given by a band index, a crystal wave vector and a spin state: $I=(i,\bm k_1,s_1), J=(j,\bm k_2,s_2)$, etc. Since the Coulomb operator does not affect the spin degree of freedom the two-electron integrals reduce to 
\begin{align}
\langle I J | A B \rangle &= \langle i\bm k_1s_1,j\bm k_2 s_2 | a\bm k_3s_3,b\bm k_4s_4 \rangle\nonumber\\
&= \langle i\bm k_1s_1,j\bm k_2s_2 | a\bm k_3s_1,b\bm k_4s_2 \rangle\delta_{s_1s_3}\delta_{s_2s_4} \;.
\end{align}
The spin-unrestricted MP2 energy per unit cell of a periodic system can thus be written as, $E^{(2)} = E^{(2)}_\text{d} + E^{(2)}_\text{x}$,
\begin{widetext}
\begin{align}
E^{(2)}_\text{d} &= \phantom{+}\frac 1 2 \frac 1 N \sumBZm{k}{1}{4} \sum_{ij}^{\text{occ.}} \sum_{ab}^{\text{virt.}} \sum_{ss'}^{\uparrow\downarrow} \frac{|\langle i\bm k_1s, j\bm k_2s' | a\bm k_3s, b\bm k_4s' \rangle |^2}{\varepsilon_{i\bm k_1s} + \varepsilon_{j\bm k_2s'} - \varepsilon_{a\bm k_3s} - \varepsilon_{ b\bm k_4s'} } \;, \label{eq:MP2direct}  \\
E^{(2)}_\text{x} &= -\frac 1 2 \frac 1 N \sumBZm{k}{1}{4} \sum_{ij}^{\text{occ.}} \sum_{ab}^{\text{virt.}} \sum_{s}^{\uparrow\downarrow} \frac{\langle i\bm k_1s, j\bm k_2s | a\bm k_3s, b\bm k_4s \rangle \langle a\bm k_3s, b\bm k_4s | j\bm k_2s, i\bm k_1s \rangle }{\varepsilon_{i\bm k_1s} + \varepsilon_{j\bm k_2s} - \varepsilon_{a\bm k_3s} - \varepsilon_{ b\bm k_4s} } \;, \label{eq:MP2exchange}
\end{align}
\end{widetext}
where BZ stands for the first Brillouin zone. For brevity, most of the following calculations are performed for the exchange term $E^{(2)}_\text{x}$ only. Furthermore the spin-restricted case is assumed, i.e. $\sum_s^{\uparrow\downarrow} \rightarrow 2$, in order to attain a compact notation:
\begin{multline}
E^{(2)}_\text{x} = \\
-\frac 1 N \smashoperator{\sum_{\bm k_1 ... \bm k_4}^{\text{BZ}}}\sum_{ij}^{\text{occ.}} \sum_{ab}^{\text{virt.}} \frac{\langle i\bm k_1, j\bm k_2 | a\bm k_3, b\bm k_4 \rangle \langle a\bm k_3, b\bm k_4 | j\bm k_2, i\bm k_1 \rangle }{\varepsilon_{i\bm k_1} + \varepsilon_{j\bm k_2} - \varepsilon_{a\bm k_3} - \varepsilon_{ b\bm k_4} } \;. \label{eq:MP2ex_spinrestr} 
\end{multline}
%A two-electron integral can be calculated in a way that scales linearly with the system size. This can be attained by an evaluation in the reciprocal space by means of a fast Fourier transform (see Eq. \ref{eq:2eIPLWV}). The fact that they need to be evaluated for all $i,j,a,b$ leads to a quintic scaling \cite{Marsman2009}. However, it is possible to perform the summation over the virtual bands $a, b$ in advance such that the scaling reduces to quartic order. Accordingly the sums over the bands need to be decoupled. This can be achieved by a Laplace transformation of the energy denominator \cite{Almlof1991}.
Note that the two-electron integrals $\langle i\bm k_1, j\bm k_2 | a\bm k_3, b\bm k_4 \rangle$ are non-vanishing only if $\bm k_1 + \bm k_2 = \bm k_3 + \bm k_4 + \bm G$, where $\bm G$ is some arbitrary reciprocal lattice vector. This can be interpreted as crystal momentum conservation. Hence, the two-electron integrals depend only on three $\bm k$-points. In Appendix \ref{A:TEFOPW} the two-electron integrals are written in the plane wave basis, revealing this crystal momentum conservation. However, beside this cubic scaling in the number of $\bm k$-points the system size scaling is still quintic. In the next step we apply the aforesaid Laplace transform in order to decouple the band summations.

\subsection{Laplace transformed MP2}

In (\ref{eq:MP2direct}), (\ref{eq:MP2exchange}), and (\ref{eq:MP2ex_spinrestr}) the sums over $i,j,a,b$ can be decoupled by applying a Laplace transformation of the energy denominator \cite{Almlof1991}:
\begin{multline}
\frac{1}{\varepsilon_{i\bm k_1} + \varepsilon_{j\bm k_2} - \varepsilon_{a\bm k_3} - \varepsilon_{ b\bm k_4}} \\
= -\int_0^\infty \D\tau\;\EuE^{(\varepsilon_{i\bm k_1} + \varepsilon_{j\bm k_2} - \varepsilon_{a\bm k_3} - \varepsilon_{ b\bm k_4})\tau} \;.
\end{multline}
Note that $\varepsilon_{i\bm k_1},\varepsilon_{j\bm k_2}<\varepsilon_\text{F}$ and $\varepsilon_{a\bm k_3},\varepsilon_{ b\bm k_4}>\varepsilon_\text{F}$ where $\varepsilon_\text{F}$ is the Fermi energy. Thus the positive definiteness of the denominator and exponent is guaranteed. This leads to the well-known Laplace transformed MP2 (LTMP2) expression \cite{Almlof1991}:
\begin{multline}
E^{(2)}_\text{x} = \frac 1 N\int_0^\infty \D\tau \sumBZm{k}{1}{4} \sum_{ij}^{\text{occ.}} \sum_{ab}^{\text{virt.}} \langle i\bm k_1, j\bm k_2 | a\bm k_3, b\bm k_4 \rangle \\
\times \langle a\bm k_3, b\bm k_4 | j\bm k_2, i\bm k_1 \rangle \; \EuE^{(\varepsilon_{i\bm k_1} + \varepsilon_{j\bm k_2} - \varepsilon_{a\bm k_3} - \varepsilon_{ b\bm k_4})\tau}  \;.  \label{eq:LTMP2}
\end{multline}
Although the decoupling comes at the price of an additional integration, it has no effect on the scaling. Due to the exponentially decreasing behavior in $\tau$, the integration can be performed by a quadrature \cite{Haser1992, Kaltak2014} using only a few $\tau$-points.\\
As already indicated in Sec. \ref{sec:brief_strategy}, the summations over the virtual bands $a,b$ can now be performed in advance. In the following we will derive an analogous decomposition to Eq. (\ref{eq:SCHnewstate}) for periodic system.

\subsection{LTMP2 in the plane wave basis and presummation over all virtual bands}

The algorithm presented here is implemented in \textsc{vasp} which uses a plane wave basis set. Hence the two-electron integrals $\langle i\bm k_1, j\bm k_2 | a\bm k_3, b\bm k_4 \rangle$ will be evaluated in the plane wave basis. The orbitals obey Bloch's theorem due to the periodicity of the system. In position representation they can be written as
\begin{equation}
\varphi_{i\bm k}(\bm r) = \langle \bm r | i \bm k \rangle  = \frac{1}{\sqrt \Omega} \EuE^{\ImI\bm k \bm r}u_{i \bm k}(\bm r) \;, \label{eq:BlochOrbital}
\end{equation}
where $u_{i \bm k}$ is the cell periodic part of the orbital. The system is decomposed into $N$ unit cells of volume $\Omega_0$ such that the volume of the entire system is $\Omega = N\Omega_0$. At the boundaries the Born-von Karman conditions are assumed. The states are normalized by 
\begin{equation}
\langle i \bm k_1 | j \bm k_2 \rangle = \int_\Omega \D^3 r  \; \varphi^*_{i\bm k_1}(\bm r) \varphi_{j\bm k_2}(\bm r) = \delta_{ij}\delta_{\bm k_1 \bm k_2} \;, \label{eq:BlochNorm}
\end{equation}
which implies
\begin{equation}
\int_{\Omega_0} \D^3 r  \; u^*_{i\bm k}(\bm r) u_{j\bm k}(\bm r) = \Omega_0 \delta_{ij} \;.
\end{equation}
%
%The two-electron integrals can be written as
%\begin{multline}
%\langle i\bm k_1, j\bm k_2 | a\bm k_3, b\bm k_4 \rangle  \\
%= \int_\Omega \D\bm r\int_\Omega \D\bm r' \frac{\varphi_{i\bm k_1}^*(\bm r) \varphi_{j\bm k_2}^*(\bm r')  \varphi_{a\bm k_3}(\bm r)  \varphi_{b\bm k_4}(\bm r')  } {|\bm r - \bm r'|}  \;. \label{eq:2eIposition}
%\end{multline}
Using Eq. (\ref{eq:BlochOrbital}) and expanding the Coulomb kernel, $1/|\bm r - \bm r'|$, in Fourier space, the two-electron integrals can be written as:
\begin{align}
& \phantom{=} \langle i\bm k_1, j\bm k_2 | a\bm k_3, b\bm k_4 \rangle \nonumber \\
& = \frac{1}{\Omega} \delta_{T(\bm k_1 - \bm k_3),T(\bm k_4 - \bm k_2)} \sum_{\bm G} \frac{4\pi}{[\bm G + T(\bm k_1 - \bm k_3)]^2} \nonumber \\
& \times \langle i\bm k_1 | \EuE^{+\ImI[\bm G + T(\bm k_1 - \bm k_3)]\hat{\bm r}} | a \bm k_3 \rangle_{\Omega_0} \nonumber \\
& \times \langle j\bm k_2 | \EuE^{-\ImI[\bm G + T(\bm k_4 - \bm k_2)]\hat{\bm r}} | b \bm k_4 \rangle_{\Omega_0} \;. \label{eq:2eIPLWV}
\end{align}
A step-by-step derivation can be found in Appendix \ref{A:TEFOPW}. An explanation of the notation is in order: $\sum_{\bm G}$ is a sum over all (infinitely many) reciprocal lattice vectors $\bm G$ arising from the real unit cell (or supercell) of the system. $T(\bm k)$\label{def:T} is a function that maps $\bm k$ back to the first BZ along a translation by an appropriate reciprocal lattice vector. In analogy to (\ref{eq:SCHovrlp}) the overlap densities are defined by
\begin{equation}
\langle i\bm k_1 | \EuE^{-\ImI\bm G\hat{\bm r}} | a \bm k_3 \rangle_{\Omega_0}  :=N\int_{\Omega_0} \D^3 r \; \varphi_{i\bm k_1}^*(\bm r) \varphi_{a\bm k_3}(\bm r) \; \EuE^{-\ImI\bm G \bm r} \;. \label{eq:OverlapDens}
\end{equation}
Note that this unitless quantity does not explicitly depend on $N$, since $N$ is balanced by the normalization factors of the orbitals (\ref{eq:BlochOrbital}). However, there is an implicit dependence since different $N$ lead to different Born-von Karman boundaries, hence a different mesh of crystal wave vectors ($\bm k$-point mesh).  Inserting (\ref{eq:2eIPLWV}) into (\ref{eq:LTMP2}), one sum over the BZ, here $\bm k_4\rightarrow T(\bm k_1 + \bm k_2 - \bm k_3)$, can be eliminated. Note that the Kronecker deltas for the $\bm k$-vectors which occur in $\langle i\bm k_1, j\bm k_2 | a\bm k_3, b\bm k_4 \rangle$ and $\langle a\bm k_3, b\bm k_4 | j\bm k_2, i\bm k_1 \rangle$ are equivalent. A substitution $\bm q = T(\bm k_2 - \bm k_3)$ then leads to
\begin{widetext}
\begin{align}
E^{(2)}_\text{x} =&  \frac{1}{\Omega_0^2}\frac{1}{N^3} \int_0^\infty \D\tau \sum_{\bm k_1 \bm k_2 \bm q}^{\text{BZ}} \sum_{ij}^{\text{occ.} } \sum_{\bm G} \frac{4\pi}{[\bm G + T(\bm k_1 - \bm k_2 + \bm q)]^2}  \sum_{\bm G'} \frac{4\pi}{(\bm G' - \bm q)^2}  \nonumber \\
& \times \sum_{a}^{\text{virt.}} \langle i\bm k_1 | \EuE^{+\ImI[\bm G + T(\bm k_1 - \bm k_2 + \bm q)]\hat{\bm r}} | a T(\bm k_2 - \bm q) \rangle_{\Omega_0}\;  \times \; { \langle a T(\bm k_2 - \bm q) | \EuE^{+\ImI(\bm G' -  \bm q)\hat{\bm r}} | j \bm k_2 \rangle_{\Omega_0} \; \EuE^{(\varepsilon_{j\bm k_2} - \varepsilon_{aT(\bm k_2 - \bm q) } )\tau}} \nonumber\\
& \times \sum_{b}^{\text{virt.}} \langle j\bm k_2 | \EuE^{-\ImI[\bm G + T(\bm k_1 - \bm k_2 + \bm q)]\hat{\bm r}} | b T(\bm k_1 + \bm q) \rangle_{\Omega_0} \; \times \; {\langle b T(\bm  k_1 + \bm q) | \EuE^{-\ImI(\bm G' -  \bm q)\hat{\bm r}} | i \bm k_1 \rangle_{\Omega_0} \; \EuE^{(\varepsilon_{i\bm k_1} - \varepsilon_{ bT(\bm k_1 + \bm q)})\tau}} \;. \label{eq:pre_xLTMP2}
%E^{(2)}_\text{x} =&  \frac{1}{\Omega_0^2}\frac{1}{N^3} \int_0^\infty \D\tau \sum_{\bm k_1 \bm k_2 \bm q}^{\text{BZ}} \sum_{ij}^{\text{occ.} } \sum_{\bm G} \frac{4\pi}{[\bm G + T(\bm k_1 - \bm k_2 + \bm q)]^2}  \sum_{\bm G'} \frac{4\pi}{(\bm G' - \bm q)^2}  \nonumber \\
%& \times \sum_{a}^{\text{virt.}} \langle i\bm k_1 | \EuE^{+\ImI[\bm G + T(\bm k_1 - \bm k_2 + \bm q)]\hat{\bm r}} | a T(\bm k_2 - \bm q) \rangle_{\Omega_0}\;  \times \; \underbracket[0.5pt]{ \langle a T(\bm k_2 - \bm q) | \EuE^{+\ImI(\bm G' -  \bm q)\hat{\bm r}} | j \bm k_2 \rangle_{\Omega_0} \; \EuE^{(\varepsilon_{j\bm k_2} - \varepsilon_{aT(\bm k_2 - \bm q) } )\tau}} \nonumber\\
%& \times \sum_{b}^{\text{virt.}} \langle j\bm k_2 | \EuE^{-\ImI[\bm G + T(\bm k_1 - \bm k_2 + \bm q)]\hat{\bm r}} | b T(\bm k_1 + \bm q) \rangle_{\Omega_0} \; \times \; \underbracket[0.5pt]{\langle b T(\bm  k_1 + \bm q) | \EuE^{-\ImI(\bm G' -  \bm q)\hat{\bm r}} | i \bm k_1 \rangle_{\Omega_0} \; \EuE^{(\varepsilon_{i\bm k_1} - \varepsilon_{ bT(\bm k_1 + \bm q)})\tau}} \;. \label{eq:pre_xLTMP2}
\end{align}
\end{widetext}
The MP2 energy per unit cell does not explicitly depend on $N$, although $1/N^3$ appears in the above formula. Again the $N$ dependence is only implicit since $N$ defines the density of the allowed $\bm k$-points. This becomes evident when we perform the thermodynamic limit, $N\rightarrow\infty$, $\Omega/N = \Omega_0 = \text{const.}$ . The density of the crystal wave vectors $\bm k$ becomes infinite and sums turn into integrals:
\begin{equation}
\frac 1 N \sum_{\bm k} ... \; \longrightarrow \; \Omega_0 \intBZpi{k} \; ... \;,
\end{equation}
which implies:
\begin{equation}
\frac{1}{\Omega_0^2} \frac{1}{N^3}\sum_{\bm k_1 \bm k_2 \bm q}^{\text{BZ}} ...  \;\longrightarrow\; \Omega_0 \intBZpiT{k_1}{k_2}{q}\,...  \;.
\end{equation}
Since $\Omega\rightarrow \infty$ the normalization factor has to be dropped in (\ref{eq:BlochOrbital}) and we write $\varphi_{i\bm k}(\bm r) \longrightarrow \EuE^{\ImI\bm k \bm r}u_{i\bm k} (\bm r)$ such that the overlap densities (\ref{eq:OverlapDens}) now read:
\begin{equation}
\langle i\bm k_1 | \EuE^{-\ImI\bm G\hat{\bm r}} | a \bm k_3 \rangle_{\Omega_0} \longrightarrow  \frac 1 \Omega_0 \int_{\Omega_0} \D^3 r \, \varphi_{i\bm k_1}^*(\bm r) \varphi_{a\bm k_3}(\bm r) \, \EuE^{-\ImI\bm G \bm r} \;. \label{eq:overlaps_thermo_limit}
\end{equation}

The formula (\ref{eq:pre_xLTMP2}) already suggests a possibility to perform a presummation over all virtual bands $a, b$: For all $\bm G'$, $\bm q$ and $\tau$  define a decomposition by 
\begin{equation}
\Big | {w}_{i\bm k}^{(\bm G'\bm q \tau)} \Big \rangle  = \sum_b^{\text{virt.}} C^{(\bm G'\bm q\tau)}_{ib,\bm k}\,|b\bm k\rangle\;. \label{eq:StateDec}
\end{equation}
The coefficients are defined as:
\begin{align}
C^{(\bm G'\bm q\tau)}_{ib,\bm k} & =  \EuE^{(\varepsilon_{iT(\bm k-\bm q)} -\varepsilon_{b\bm k})\tau} \nonumber\\
& \phantom{=} \times  \langle b \bm  k | \EuE^{-\ImI(\bm G' -  \bm q)\hat{\bm r}} | i T(\bm k - \bm q) \rangle_{\Omega_0} \;. \label{eq:DecompCoeff}
\end{align}
The decomposition (\ref{eq:StateDec}) can be performed in advance such that for a given $\bm G'$, $\bm q$ and $\tau$ the large number of virtual bands are reduced to transformed states $\big|w_{i\bm k}^{(\bm G'\bm q \tau)}\big\rangle$ labeled by the few occupied indices $i$. Hence the MP2 energy per unit cell can be written in a form which involves summations only over the occupied indices $i,j$:
%\begin{widetext}
\begin{align}
%\begin{multline}
%E^{(2)}_\text{x} &= \Omega_0 \int_0^\infty \D\tau \intBZpiT{q}{k_1}{k_2}   \sum_{\bm G\bm G'} \frac{4\pi}{[\bm G + T(\bm k_1 - \bm k_2 + \bm q)]^2}  \frac{4\pi}{(\bm G' - \bm q)^2}  \nonumber \\
%&\phantom{=}\times \sum_{ij}^{\text{occ.} } \Big\langle i\bm k_1 \Big| \EuE^{+\ImI[\bm G + T(\bm k_1 - \bm k_2 + \bm q)]\hat{\bm r}} \Big| {w}_{jT(\bm k_2 - \bm q)}^{(-\bm G',-\bm q, \tau)} \Big\rangle_{\Omega_0}\;  \Big\langle j\bm k_2 \Big| \EuE^{-\ImI[\bm G + T(\bm k_1 - \bm k_2 + \bm q)]\hat{\bm r}} \Big| {w}_{iT(\bm k_1+\bm q)}^{(\bm G'\bm q \tau)}  \Big\rangle_{\Omega_0} \;. \label{eq:xLTMP2}
E^{(2)}_\text{x}  = \Omega_0 & \int_0^\infty \D\tau \intBZpiT{q}{k_1}{k_2} \label{eq:xLTMP2} \\
\times \; \sum_{\bm G\bm G'} &\frac{4\pi}{[\bm G + T(\bm k_1 - \bm k_2 + \bm q)]^2}  \frac{4\pi}{(\bm G' - \bm q)^2}   \nonumber \\
\times \;\; \sum_{ij}^{\text{occ.} } &\Big\langle i\bm k_1 \Big| \EuE^{+\ImI[\bm G + T(\bm k_1 - \bm k_2 + \bm q)]\hat{\bm r}} \Big| {w}_{jT(\bm k_2 - \bm q)}^{(-\bm G',-\bm q, \tau)} \Big\rangle_{\Omega_0} \nonumber \\
\times \qquad &\Big\langle j\bm k_2 \Big| \EuE^{-\ImI[\bm G + T(\bm k_1 - \bm k_2 + \bm q)]\hat{\bm r}} \Big| {w}_{iT(\bm k_1+\bm q)}^{(+\bm G',+\bm q \tau)}  \Big\rangle_{\Omega_0} \;. \nonumber \label{eq:xLTMP2}
\end{align}
%\end{multline}
%\end{widetext}
%The derivation for the direct term (\ref{eq:dLTMP2}) is straight forward.  

\subsection{Time reversal symmetry}

For a more convenient implementation in a computer code it is advantageous to exploit the time reversal symmetry. With its aid we can turn both overlap densities in Eq. (\ref{eq:xLTMP2}) into the same form, i.e. to avoid mixtures of $+\bm G$ and $-\bm G$ as well as mixtures of $+\bm G'$ and $-\bm G'$. This can be done in the following way. If no external magnetic field is applied and spin-orbit coupling is ignored the Hartree-Fock orbitals (\ref{eq:BlochOrbital}) obey time reversal symmetry: $\varphi_{i\bm k}^* = \varphi_{i-\bm k}$ and $\varepsilon_{i\bm k} = \varepsilon_{i-\bm k}$. If we apply this time reversal symmetry to the coefficients (\ref{eq:DecompCoeff}) we obtain:
\begin{equation}
\left( C^{(+\bm G',+\bm q,\tau)}_{ib,+\bm k} \right)^* = C^{(-\bm G',-\bm q,\tau)}_{ib,-\bm k} \;.
\end{equation}
Hence for (\ref{eq:StateDec}) we find:
\begin{equation}
\Big | \Big( {w}_{i,+\bm k}^{(+\bm G',+\bm q, \tau)} \Big)^* \Big \rangle = \Big | {w}_{i,-\bm k}^{(-\bm G',-\bm q, \tau)} \Big \rangle \;. \label{eq:DecStateSym}
\end{equation}
This relation can be applied to the overlap densities in (\ref{eq:xLTMP2}). Consider, e.g.,
\begin{align}
& \Big\langle i\bm k_1 \Big| \EuE^{+\ImI[\bm G + T(\bm k_1 - \bm k_2 + \bm q)]\hat{\bm r}} \Big| {w}_{j T(\bm k_2 - \bm q)}^{(-\bm G',-\bm q, \tau)} \Big\rangle_{\Omega_0}  \nonumber \\
=\; & \Big\langle {w}_{j T(\bm k_2 - \bm q)}^{(-\bm G',-\bm q, \tau)} \Big| \EuE^{-\ImI[\bm G + T(\bm k_1 - \bm k_2 + \bm q)]\hat{\bm r}} \Big|  i\bm k_1  \Big\rangle_{\Omega_0}^* \\
= \; & \Big\langle (i\bm k_1)^* \Big| \EuE^{-\ImI[\bm G + T(\bm k_1 - \bm k_2 + \bm q)]\hat{\bm r}} \Big| {w}_{j T(-\bm k_2 + \bm q)}^{(\bm G'\bm q \tau)} \Big\rangle_{\Omega_0}^*  \nonumber
\end{align}
%& \intBZpi{k_1} \Big\langle i\bm k_1 \Big| \EuE^{+\ImI(\bm G + \bm q)\hat{\bm r}} \Big| {w}_{iT(\bm k_1 - \bm q)}^{(-\bm G',-\bm q, \tau)} \Big\rangle_{\Omega_0} \nonumber\\
%=& \intBZpi{k_1}\Big\langle  {w}_{iT(\bm k_1 - \bm q)}^{(-\bm G',-\bm q,\tau)} \Big| \EuE^{-\ImI(\bm G + \bm q)\hat{\bm r}} \Big| i\bm k_1 \Big\rangle_{\Omega_0}^* \nonumber \\
%=& \intBZpi{k_1}\Big\langle i\bm k_1 \Big| \EuE^{-\ImI(\bm G + \bm q)\hat{\bm r}} \Big| {w}_{iT(\bm k_1 + \bm q)}^{(\bm G'\bm q \tau)} \Big\rangle_{\Omega_0}^* \;.
In a last step we substitute $\bm k_2 \rightarrow -\bm k_2$ in (\ref{eq:xLTMP2}) such that we find a convenient and more ''symmetric'' formula for the MP2 exchange energy per unit cell:
%\begin{widetext}
\begin{align}
E^{(2)}_\text{x} = \Omega_0 & \int_0^\infty \D\tau \intBZpiT{q}{k_1}{k_2} \label{eq:xLTMP2Compact}\\
\times \; \sum_{\bm G\bm G'} & \frac{4\pi}{[\bm G + T(\bm k_1 + \bm k_2 + \bm q)]^2}  \frac{4\pi}{(\bm G' - \bm q)^2}  \nonumber \\
\times \;\; \sum_{ij}^{\text{occ.} } & \Big\langle (i\bm k_1)^* \Big| \EuE^{-\ImI[\bm G + T(\bm k_1 + \bm k_2 + \bm q)]\hat{\bm r}} \Big| {w}_{jT(\bm k_2 + \bm q)}^{(\bm G'\bm q, \tau)} \Big\rangle_{\Omega_0}^* \nonumber \\
\times \qquad & \Big\langle (j\bm k_2)^* \Big| \EuE^{-\ImI[\bm G + T(\bm k_1 + \bm k_2 + \bm q)]\hat{\bm r}} \Big| {w}_{iT(\bm k_1+\bm q)}^{(\bm G'\bm q \tau)}  \Big\rangle_{\Omega_0} \;. \nonumber 
\end{align}
%\end{widetext}
As for the exchange term the same procedure can be applied to the direct term. Starting with (\ref{eq:MP2direct}) one finds: 
\begin{multline}
E^{(2)}_\text{d} = -2 \Omega_0 \int_0^\infty \D\tau \intBZpi{q} \sum_{\bm G\bm G'} \frac{4\pi}{(\bm G + \bm q)^2}   \frac{4\pi}{(\bm G' - \bm q)^2} \\
\times \left| \; \intBZpi{k}  \sum_{i}^{\text{occ.} } \Big\langle i\bm k \Big| \EuE^{-\ImI(\bm G + \bm q)\hat{\bm r}} \Big| {w}_{iT(\bm k + \bm q)}^{(\bm G'\bm q \tau)} \Big\rangle_{\Omega_0}\right|^2 \;. \label{eq:dLTMP2Compact}  
\end{multline}
%The scaling of the direct term (\ref{eq:dLTMP2Compact}) is cubic in the system size (sum over $\bm G, \bm G', i$) and quadratic in the number of k-points (sum over $\bm q, \bm k$). For the exchange term (\ref{eq:xLTMP2Compact}) the scaling is quartic in the system size (sum over $\bm G, \bm G', i, j$) and cubic in the number of k-points (sum over $\bm k_1, \bm k_2, \bm q$). The number of $\tau$-points of the quadrature, to numerically evaluate the $\tau$-integration, is independent of the system size. 

%%%%%%%%%%%%%%%%%%%%%%%%%%%%%%%%%%%%%%%%%%%%%%%%%%%%%%%%%%%%%%%%%
\section{Implementation}\label{sec:Impl}
%%%%%%%%%%%%%%%%%%%%%%%%%%%%%%%%%%%%%%%%%%%%%%%%%%%%%%%%%%%%%%%%%

The presented LTMP2 method is implemented in the Vienna ab-initio simulation package (\textsc{vasp}) \cite{Kresse1993,Kresse1999} based on the projector augmented wave (PAW) \cite{Blochl1994} method. For brevity this section restricts to the $\bm \Gamma$-only version, i.e. we neglect $\bm k$-point sampling of the Brillouin zone here. 

\subsection{General strategy}

The four major steps of the algorithm can be summarized as follows:\\
%(I) Calculate and store all overlap densities (\ref{eq:overlaps_thermo_limit}).\\
%(II) Loop over all $\tau$-points and plane waves $\bm G$.\\
%(III) Calculate the transformation matrix (\ref{eq:DecompCoeff}) using the stored overlap densities and construct the transformed states (\ref{eq:StateDec}).
\label{item:stepii}
\begin{enumerate}[(i)]
\item Compute and store all overlap densities (\ref{eq:SCHovrlp}).
\item Loop over all $\tau$-points and reciprocal vectors $\bm G$ of the outer loops in Eq. (\ref{eq:SCHfocklike}).
\item  Calculate the transformation matrix (\ref{eq:SCHtransmatrix}) for this $\tau$ and $\bm G$ using the stored overlap densities of step (i)  and construct the transformed states $|w_i\rangle$, see (\ref{eq:SCHnewstate}).
\item Perform
\begin{enumerate}[(a)]
\item a Hartree-like calculation,
\item a Fock-like calculation,
\end{enumerate}
to calculate the direct and exchange MP2 contribution for this $\tau$ and $\bm G$, i.e. evaluate the two-electron integral $\langle ij |  w_j  w_i \rangle$ and sum over $i,j$ in (\ref{eq:SCHfocklike}).
\end{enumerate}

\begin{figure}

\includegraphics[width=0.8\linewidth]{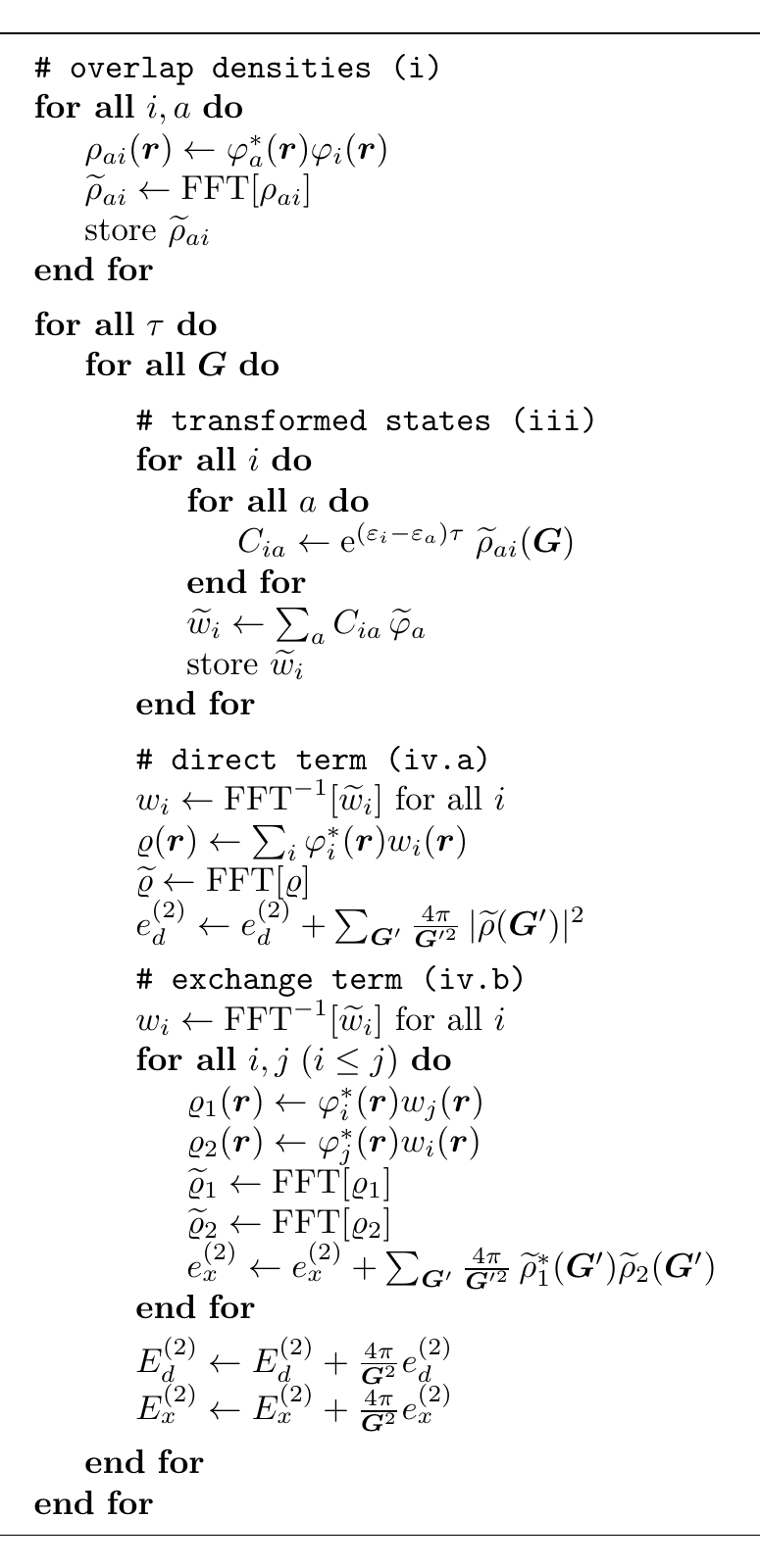}

\caption{Pseudocode of the serial $\bm \Gamma$-only implementation. Hartree-Fock orbitals $\varphi_i,\varphi_a$ and energies $\varepsilon_i,\varepsilon_a$ for all occupied and virtual states are assumed. }
\label{fig:SerialGamma}
\end{figure}

The outer loop over plane waves $\bm G$ is limited by an adjustable plane wave cutoff. The outer loop over the $\tau$-points is performed by a quadrature \cite{Kaltak2014}. Figure \ref{fig:SerialGamma} shows the pseudocode for the serial $\bm \Gamma$-only implementation of the algorithm.

\subsection{Parallelization}

The outer $\bm G$-loop (see step (ii) in Sec. \ref{item:stepii} or Eq. \ref{eq:SCHfocklike}) provides a powerful approach to parallelize the algorithm. The entire set of reciprocal lattice vectors $\mathcal G$ can be divided into $\mathcal N_{\bm G}$ independent subsets. This leads to a very high parallelization efficiency as long as the total number of reciprocal lattice vectors $N_{\bm G}$ is larger than $\mathcal N_{\bm G}$. Furthermore, on a second level, a parallelization is implemented for the evaluation of the sum over occupied bands: the set of occupied bands $\mathcal I$  is divided into $\mathcal N_\text{B}$ subsets such that one band summation (say over $j$) can be calculated in parallel. Pseudocode for this strategy can be found in Fig. \ref{fig:ParallelGamma}.

\subsection{Formal system size scaling:\\computation time and memory}

%spatial integrals -> spatial summation (FFT grid)
%tau integrals -> quadrature (Merzuk)
%scaling behavior here and not in the theory part

\begin{table}
\revcolI{
\begin{tabular}{ccc}
step   & \quad system size \quad &  k-points   \\
(i)    & $N_i N_a N_{\texttt{FFT}}\ln N_{\texttt{FFT}}$ & $N_{\bm k}^2$ \\
(iii)  & $N_i N_a N_{\bm G}^2$ & $N_{\bm k}^2$\\
(iv.a) & $N_iN_{\bm G}N_\texttt{FFT}\ln N_{\texttt{FFT}}$ & $N_k^2$\\
(iv.b) & $N_i^2N_{\bm G}N_\texttt{FFT}\ln N_{\texttt{FFT}}$ & $N_{\bm k}^3$
\end{tabular}
}
\caption{Formal scaling of the computation time for the different steps of the algorithm. The steps are described in Sec. \ref{item:stepii}. $N_i$ and $N_a$ are the number of occupied and virtual orbitals respectively. $N_{\bm G}$ is the number of reciprocal lattice vectors, $N_{\texttt{FFT}}$ the number of \texttt{FFT} grid points, and $N_{\bm k}$ is the number of $\bm k$-points.} 
\label{tab:scaling}
\end{table}

Calculating the exchange term (which has the steepest scaling) results in a computation time that scales with the fourth power of the system size. This stems from the fact that for every combination of $\bm G, i, j$ fast Fourier transforms (\texttt{FFT}) have to be performed. In a spin-unrestricted calculation, the computation time will be twice as large as for a spin-restricted case. The number of $\tau$-points for the quadrature, to numerically evaluate the $\tau$-integration, is largely independent of the system size. Table \ref{tab:scaling} shows the formal scaling for each step.

Regarding the memory consumption only the orbitals and the overlap densities (\ref{eq:SCHovrlp})  are of relevance. The latter has the steepest scaling: $N_i N_a N_{\bm G}$. The memory requirement of the orbitals scales only quadratically. Regarding $\bm k$-point sampling of the Brillouin zone, a linear dependence of $N_{\bm k}$ has to be multiplied in both cases.  For a spin-unrestricted calculation the memory requirement doubles.

\subsection{Internal cutoff extrapolation}

\label{ch:CutoffExtrapol}
Similar to other MP2 implementations or RPA codes the LTMP2 algorithm converges slowly with respect to the number of basis functions (plane wave cutoff). \revcolI{However, within the presented algorithm an internal extrapolation of the auxiliary plane wave cutoff, $E^\text{aux}_\text{cut}$, can be implemented comfortably with negligible influence on the computation time. The cutoff extrapolation makes use of the known asymptotic behavior of the MP2 energy $E^{(2)}$ for large cutoffs $E^\text{aux}_\text{cut}$ \cite{Harl2008,Shepherd2012}:
\begin{equation}
E^{(2)}(E^\text{aux}_\text{cut}) - E^{(2)}(E^\text{aux}_\text{cut} = \infty) \sim {E^\text{aux}_\text{cut}}^{-3/2} \;. \label{eq:CutDecay}
\end{equation}
The idea is to calculate the MP2 energy for a sufficient number of different cutoffs $E^\text{aux}_\text{cut}$ in order to extrapolate these energies according to Eq. (\ref{eq:CutDecay}). Since the plane wave cutoff, $E^\text{aux}_\text{cut}$, truncates only the $\bm G$-loops, the MP2 energy can be calculated for different cutoffs on the fly. This is achieved in the following way. First an array of different cutoffs is defined by
\begin{equation}
E^\text{aux}_\text{cut}\texttt{[n]} = \frac{E^\text{aux}_\text{cut}}{\alpha^{\texttt n-1}} \;,
\end{equation}
where $E^\text{aux}_\text{cut}$ is the user-given cutoff, $\alpha > 1$, $n=1,...,n_{\text c}$, and $n_{\text c}$ is the number of cutoffs for the extrapolation. In this work we chose $\alpha = 1.05$ and $n_{\text c} = 8$. Each cutoff, $E^\text{aux}_\text{cut}\texttt{[n]}$, defines the maximum length of the plane wave vectors,
\begin{equation}
G_\text{max}\texttt{[n]} = \sqrt{2E^\text{aux}_\text{cut}\texttt{[n]}}\;.
\end{equation}
Also for the MP2 energy an array, $E^{(2)}\texttt{[n]}$, is created. Whenever a loop over plane waves in the auxiliary basis, $\sum_{\bm G}...$ , is performed, only $|\bm G| \leq G_\text{max}\texttt{[n]}$ contributes to $E^{(2)}\texttt{[n]}$. This is implemented by simple \texttt{if...then...else...} statements in plane wave loops in the auxiliary basis.The different MP2 energies, $E^{(2)}\texttt{[n]}$, can then be extrapolated to $E^\text{aux}_\text{cut} \rightarrow \infty$ by solving the linear regression $y = a + b x$ with the samples $y=E^{(2)}\texttt{[n]}$ and $x=E^\text{aux}_\text{cut}\texttt{[n]}^{-3/2}$, according to Eq. (\ref{eq:CutDecay}). The solution for $a$ is the extrapolated energy $E^{(2)}(E^\text{aux}_\text{cut} = \infty)$. The reliability of this extrapolation depends on the user-given $E^\text{aux}_\text{cut}$ since the asymptotic behavior (\ref{eq:CutDecay}) is strictly true only for $E^\text{aux}_\text{cut} \rightarrow \infty$.}

%Again, as for the parallelization, the outer $\bm G$-loop can be exploited. Based on the user-given cutoff $E_\text{cut}$ which defines the largest plane wave vector $G_{\text{max}} = \sqrt{2 E_\text{cut}}$, $n_c$ smaller cutoffs are defined via $E^{(n)}_\text{cut} := E_\text{cut} \cdot 1.05^{-n}$ implying $G_{\text{max}}^{(n)} = G_{\text{max}} \cdot 1.05^{-n/2}$, $n=1,...,8$. 

%MP2 energies are computed for different cutoff spheres by calculating intermediate results of the outer $\bm G$-loop on the fly. In this way a certain number of MP2 energies are calculated for cutoffs reaching from 70\% to 100\% of the user-given plane wave cutoff $E_\text{cut}$. These MP2 energies are then extrapolated using the known asymptotic behavior for large $E_\text{cut}$ \cite{Harl2008,Shepherd2012}:
%\begin{equation}
%E^{(2)}(E_\text{cut}) - E^{(2)}(\infty) \sim E_\text{cut}^{-3/2} \;. \label{eq:CutDecay}
%\end{equation}
%The reliability of this extrapolation depends on the user-given $E_\text{cut}$ since the asymptotic behavior is strictly true only for $E_\text{cut} \rightarrow \infty$.

%%%%%%%%%%%%%%%%%%%%%%%%%%%%%%%%%%%%%%%%%%%%%%%%%%%%%%%%%%%%%%%%%
\section{Benchmark calculations}\label{sec:Bench}
%%%%%%%%%%%%%%%%%%%%%%%%%%%%%%%%%%%%%%%%%%%%%%%%%%%%%%%%%%%%%%%%%

In order to show the potential of the new LTMP2 method we performed several benchmark calculations. Computations on supercells of solid lithium hydride (LiH) served as a benchmark to demonstrate the parallelization efficiency and system size scaling. The advantage of an internal auxiliary plane wave cutoff extrapolation is shown by means of binding energy calculations of methane (CH${}_4$) in a chabazite crystal ($\text{Al}{\kern 0.1em}\text{H}{\kern 0.1em}\text{O}_{24}{\kern 0.1em}\text{Si}_{11} $). The calculations were performed with \textsc{vasp} in which the new LTMP2 code was implemented. For each benchmark calculation shown in this chapter we used a $\bm \Gamma$-only and spin-restricted setting. \revcolI{Furthermore, we used the previous MP2 implementation as a reference for the LTMP2 implementation of this work whenever possible. For the $\tau$-integration, 6 $\tau$-points turned out to be accurate enough ($\ll 1 \text{meV}$ agreement between MP2 and LTMP2) for all considered systems in this section.  }

%\begin{figure}
%\includegraphics[width=1.0\linewidth]{Chabazite_engutgw.pdf}
%\end{figure}
%\begin{figure}
%\includegraphics[width=1.0\linewidth]{NH3_3x3x3.eps}
%\end{figure}
%\begin{figure}
%\includegraphics[width=1.0\linewidth]{Chabazite.eps}
%\end{figure}

\subsection{Measured parallelization efficiency}

\begin{figure}
\includegraphics[width=1.0\linewidth]{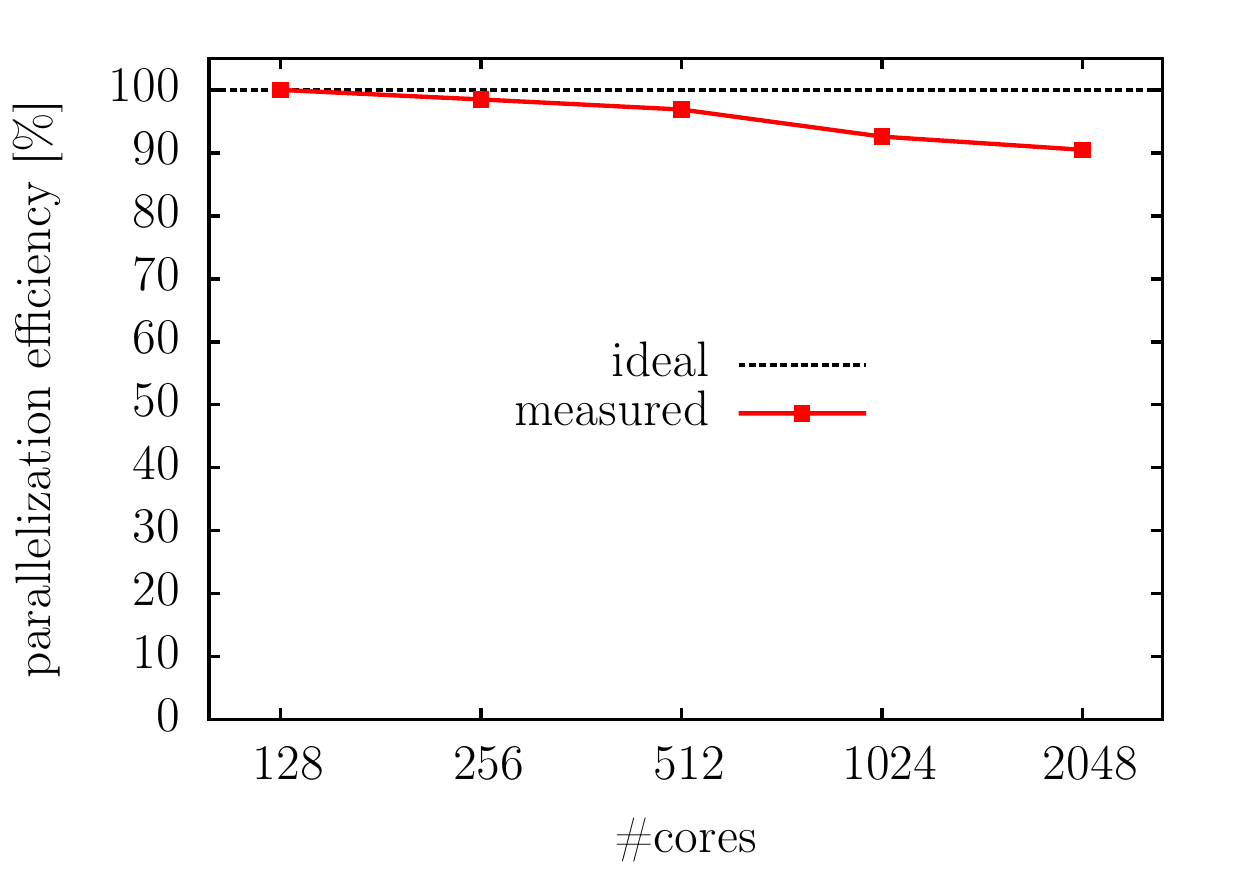}
\caption{Parallelization efficiency of the LTMP2 calculation of a $4\times4\times4$ supercell of solid LiH. The efficiency was calculated with reference to the computation time $t_{128}$ of a run using 128 cores: $t_{128}/t_{\text{\#cores}}/(\text{\#cores}/128)$. Note that the abscissa has a logarithmic scale. }
\label{fig:ParallEffic}
\end{figure}

\begin{figure}
\includegraphics[width=1.0\linewidth]{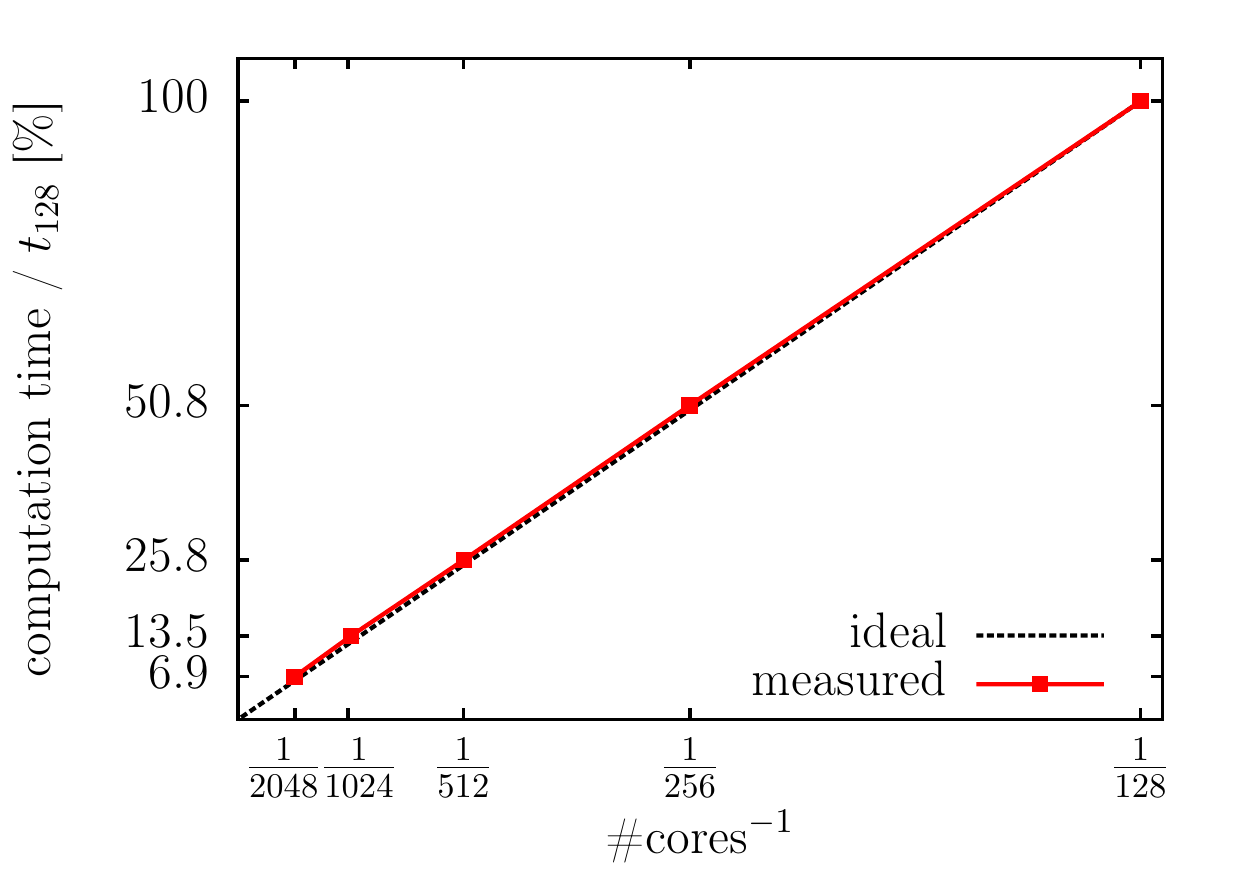}
\caption{Strong scaling of the LTMP2 calculation of a 4x4x4 supercell of solid LiH. Here $t_{128}$ is the computation time of a calculation with 128 cores. In the case of 2048 cores the computation time is reduced to 6.9\% of $t_{128}$ which is close to the ideal case of $128/2048=6.25\%$.   }
\label{fig:DeepParallScale}
\end{figure}

To demonstrate the parallelization efficiency we computed the MP2 energy of solid LiH using a supercell containing $4 \times 4 \times 4 $ primitive cells and 128 atoms. The computation time was measured against the number of cores. An auxiliary plane wave cutoff of $289\,\text{eV}$ lead to $5600$ reciprocal lattice vectors for the outer $\bm G$-loop. The total number of orbitals was set to 20480 whereas the number of occupied orbitals amounted to 128. In each run the number of plane wave groups was set to $\mathcal N_{\bm G} = \text{\#cores}/4$ which implies $\mathcal N_{\text B} = 4$ for the band parallelization.  Figures \ref{fig:ParallEffic} and \ref{fig:DeepParallScale} show the parallelization efficiency and the strong scaling. The strong scaling shows that with increasing number of cores the computation time approaches zero as long as the number  of reciprocal lattice vectors, $N_{\bm G}$, is divisible by the number of plane wave groups, $\mathcal N_{\bm G}$.

\subsection{Measured system size scaling:\qquad\qquad computation time and memory}

\begin{figure}
\includegraphics[width=1.0\linewidth]{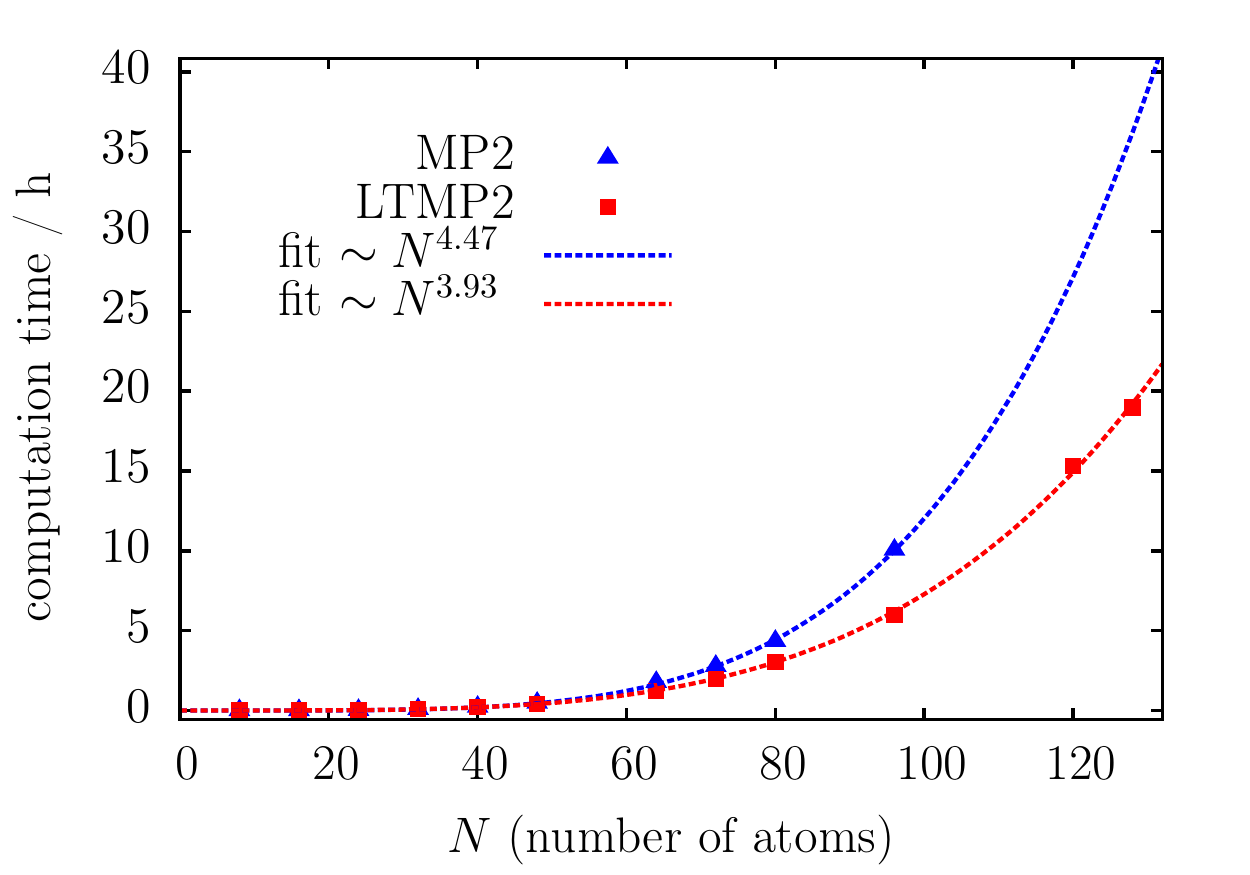}
\caption{\revcolI{System size scaling of the computation time for various supercells of solid LiH on a system with 64 cores and 8 GB memory per core. MP2 refers to the previous implementation, whereas LTMP2 is the algorithm of this work. With the previous implementation it was not possible to exceed 96 atoms due to larger memory requirements.}}
\label{fig:SysScale}
\end{figure}

\begin{figure}
\includegraphics[width=1.0\linewidth]{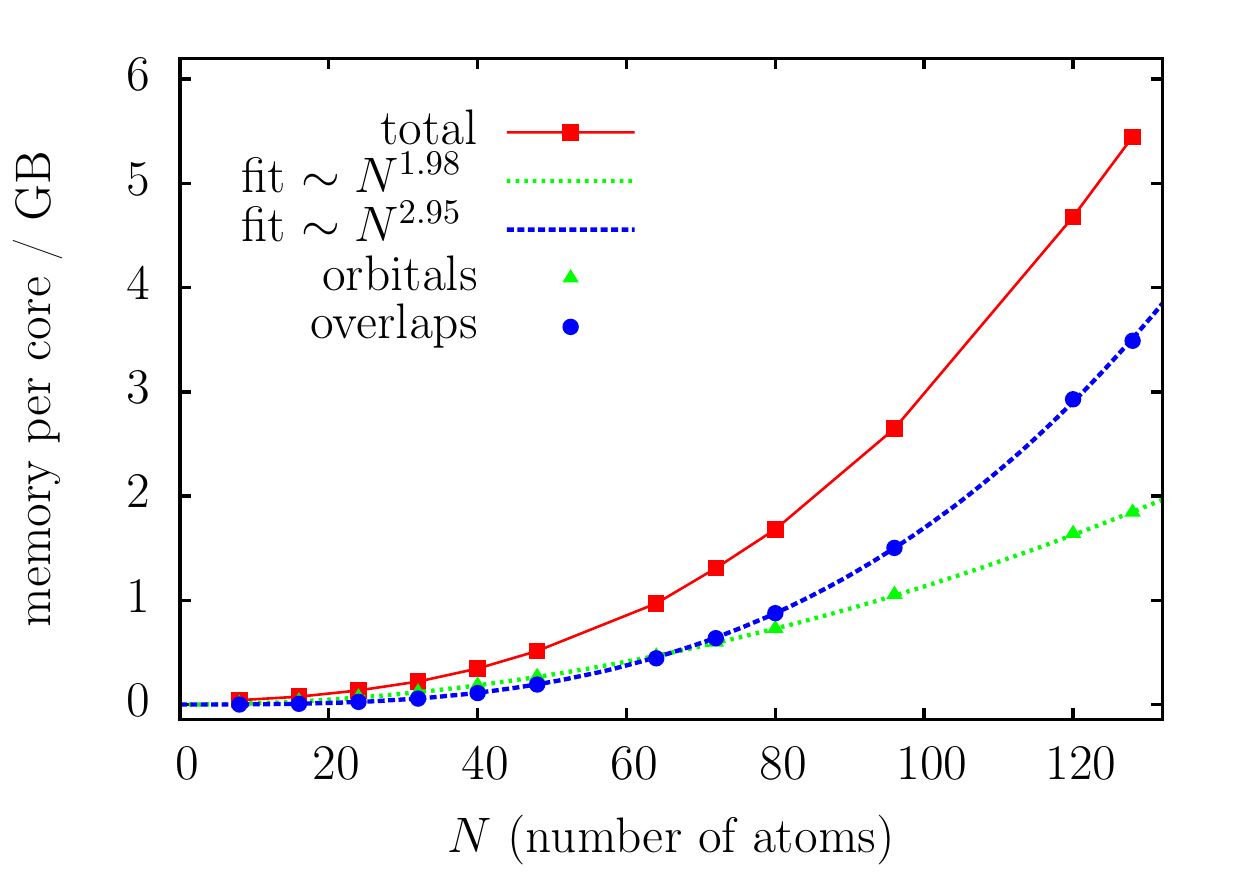}
\caption{Scaling of the used memory per core with respect to the system size. The two most relevant quantities are the orbitals and the overlap densities (\ref{eq:SCHovrlp}). \revcolI{The calculations were performed on a system with 64 cores and 8 GB memory per core.} }
\label{fig:MemScale}
\end{figure}

To evaluate the time and memory scaling of the LTMP2 algorithm with respect to the system size, the computation time and the memory consumption was measured  against the number of atoms for various supercells of solid LiH. The smallest system containing 8 atoms corresponds to the conventional unit cell of LiH. This cell was replicated in numerous ways to form supercells containing up to 128 atoms. \revcolI{The plane wave cutoff was set to 289 eV leading to about 44 plane waves of the auxiliary basis set per LiH formula. The calculations were performed on 64 Intel Xeon E5-2650 v2 2.8 GHz processors with 8 GB memory per core, although 6 GB per core would suffice for up to 128 atoms.} For the parallelization we used $\mathcal N_{\bm G} = 32$ and $\mathcal N_\text{B} = 2$. \revcolI{Figure \ref{fig:SysScale} and \ref{fig:MemScale} show the measured scaling of the computation time and memory for this (LTMP2) and the previous (MP2) implementation.} The measured scaling exponents match the predicted values.

%Regarding the computation time the measured scaling is, as might be expected, slightly below a quartic scaling. This is due to the fact that even for for 8 to 128 atoms also the cubic scaling direct MP2 contribution plays a role in the total computation time.

%The scaling of the LTMP2 algorithm with respect to the system size was tested by means of various $n \times n \times n$ supercells where $n$ goes from $1$ to $5$ containing $2$ to $250$ atoms.

\subsection{Internal cutoff extrapolation}

\begin{figure}
\includegraphics[width=1.0\linewidth]{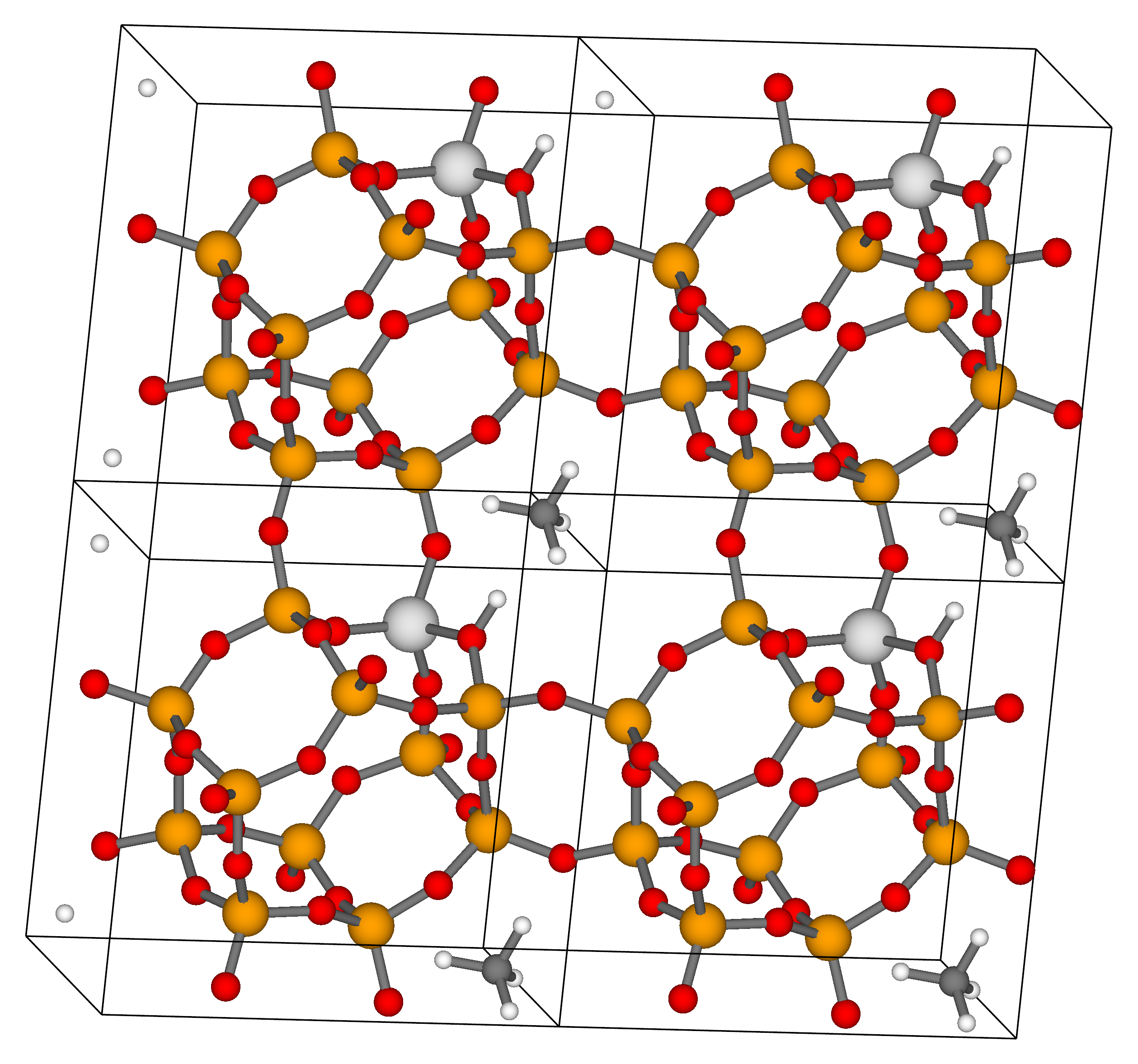}
\caption{Four unit cells of the chabazite crystal with adsorbed methane molecules. The color code reads: Al (light gray), C (dark gray), H (white), O (red), Si (yellow).}
\label{fig:Chabazite}
\end{figure}

\begin{figure}
\includegraphics[width=1.0\linewidth]{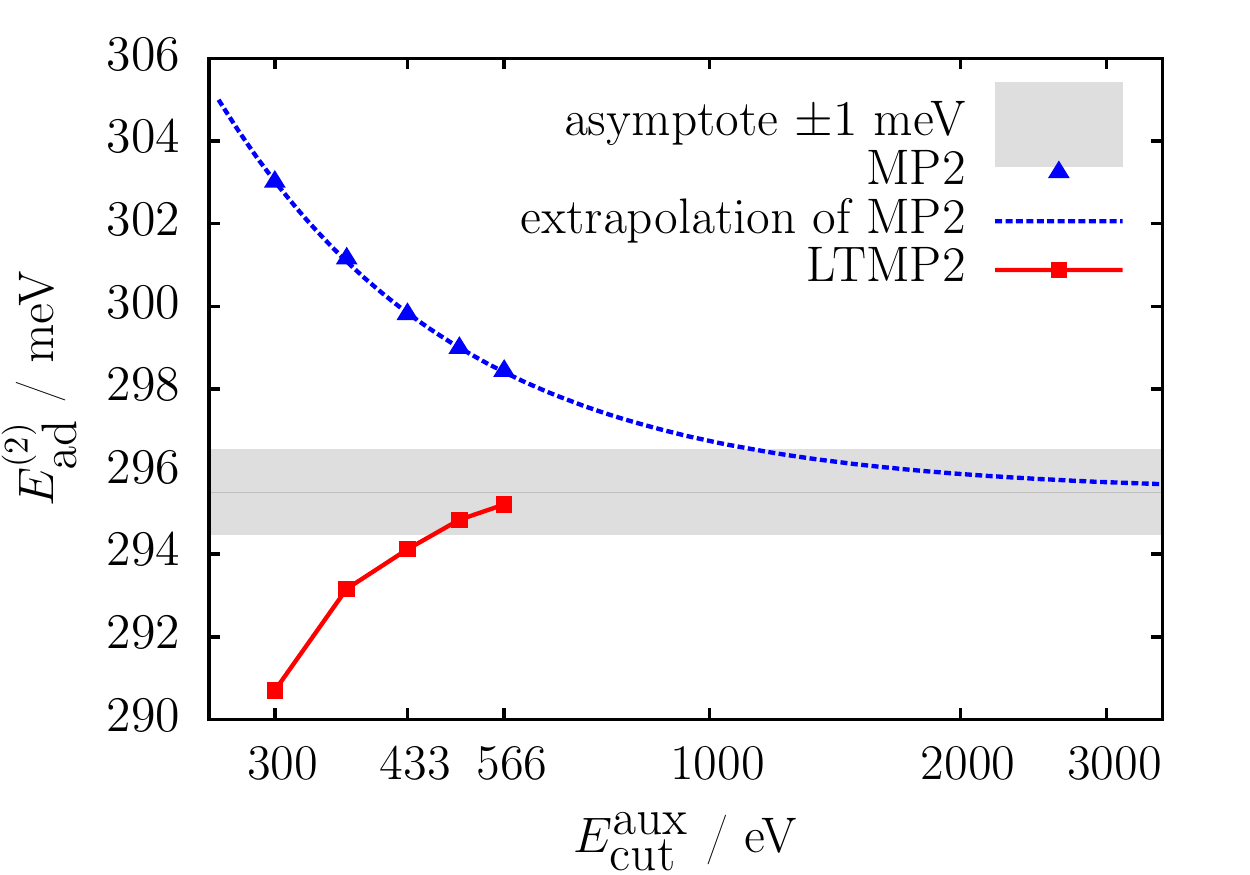}
\caption{\revcolI{Adsorption energy of methane in a chabazite cage as a function of the auxiliary plane wave cutoff $E^{\text{aux}}_{\text{cut}}$. MP2 refers to the previous implementation without basis set extrapolation, whereas LTMP2 is the algorithm of this work. The manual extrapolation (dashed blue line) was calculated using (\ref{eq:CutDecay}). Note the logarithmic scaling of the abscissa.} }
\label{fig:CutoffExtrapo}
\end{figure}

\begin{figure}
\includegraphics[width=1.0\linewidth]{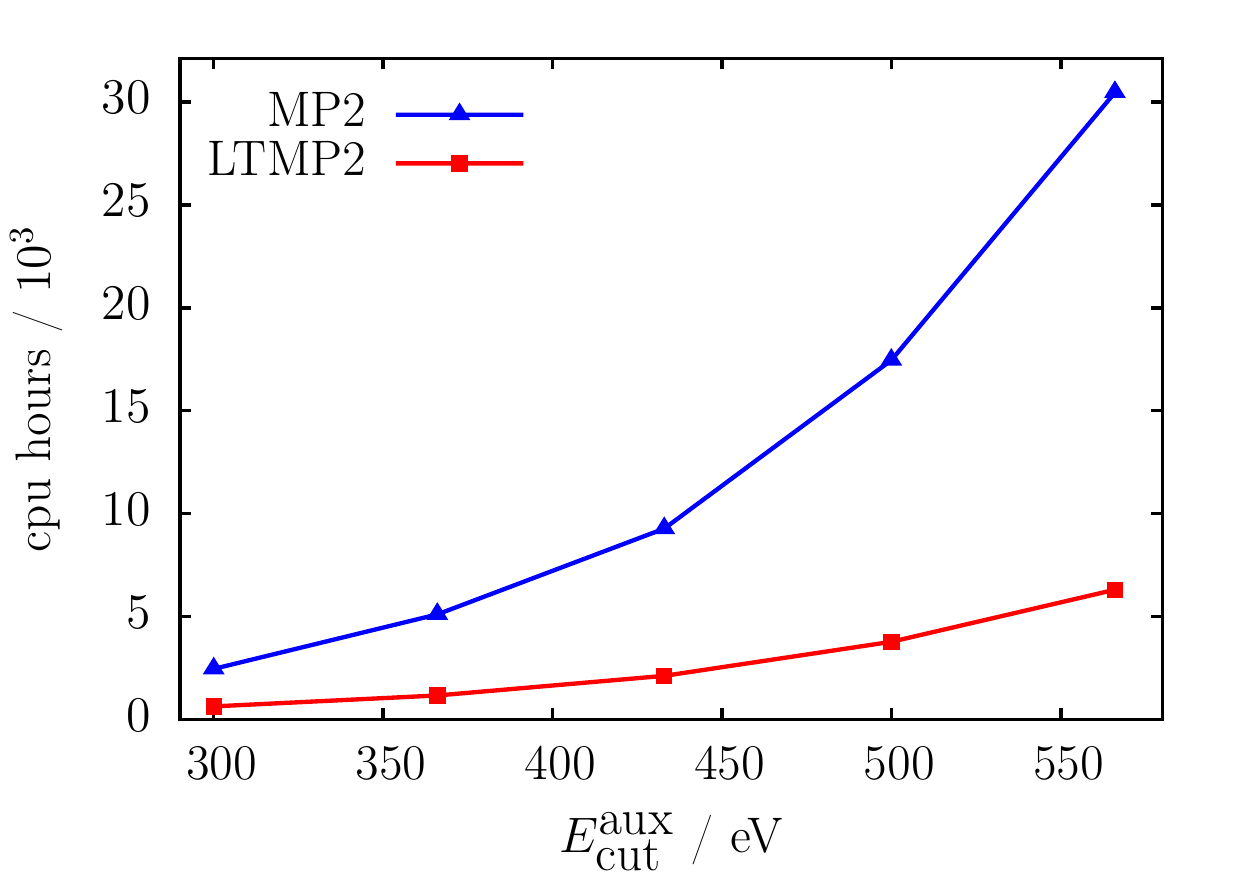}
\caption{\revcolI{CPU hours spend to calculate the adsorption energy of methane in a chabazite cage against the auxiliary plane wave cutoff $E^{\text{aux}}_{\text{cut}}$. The calculations were performed on 256 (lowest cutoff) to 1280 (highest cutoff) Intel Xeon E5-2650 v2 2.8 GHz processors with 4 GB memory per core.}}
\label{fig:CutoffTiming}
\end{figure}

\revcolI{The internal cutoff extrapolation described in Sec. \ref{ch:CutoffExtrapol} is illustrated by an adsorption energy calculation of $\text{CH}_4$ in a chabazite crystal (Chab): 
\begin{equation}
E_\text{ad}^{(2)} = E_{\text{CH}_4}^{(2)} + E_{\text{Chab}}^{(2)} - E_{\text{CH}_4+\text{Chab}}^{(2)}\;.
\end{equation}
The MP2 correlation contribution, $E_\text{ad}^{(2)}$, of the adsorption energy was computed against the auxiliary plane wave cutoff $E^\text{aux}_{\text{cut}}$. Figure \ref{fig:CutoffExtrapo} shows the result of the previous code (MP2) and this work (LTMP2). The lattice parameters and the orientation of the $\text{CH}_4$ molecule in the chabazite cage were taken from  \cite{Goltl2012} and then reoptimized using the optB88-vdW functional \cite{Klimes2010}. The largest system (CH${}_4$+Chab) consists of 42 atoms in a unit cell of about 810 $\text{\AA}^3$. A visualization can be found in Fig. \ref{fig:Chabazite}. In the Hartree-Fock steps to calculate the orbitals, plane wave cutoffs ($E_{\text{cut}}$ or \texttt{ENCUT} flag in \textsc{vasp}) of $450$, $550$, $650$, $750$, and $850$ eV were used. In the subsequent MP2 calculation auxiliary plane wave cutoffs ($E^{\text{aux}}_{\text{cut}}$ or \texttt{ENCUTGW} flag in \textsc{vasp}) of $300$, $366$, $433$, $500$, and $566$ eV were used for the two-electron integrals. In the case of $E^{\text{aux}}_{\text{cut}}=566\,\text{eV}$ the computation of $E_{\text{CH}_4+\text{Chab}}^{(2)}$ included about $46\,000$ bands ($100$ occupied bands) and about $N_{\bm G} = 12\,500$ plane wave vectors. The previous MP2 code clearly shows the mentioned ${E^{\text{aux}}_{\text{cut}}}^{-3/2}$ behavior (\ref{eq:CutDecay}) as can be seen in Fig. \ref{fig:CutoffExtrapo}. With the previous MP2 code, the procedure was to perform the extrapolation manually leading to $E_\text{ad}^{(2)} = 295.51\pm0.07\,\text{meV}$, where the error stems from the fitting. This is at the cost of about $65\,000$ cpu hours, as can be seen in Fig. \ref{fig:CutoffTiming}. Without an extrapolation this adsorption energy could not be achieved below a cutoff of $E^{\text{aux}}_\text{cut} = 1100\,\text{eV}$, if a tolerance of $1\,\text{meV}$ is assumed. However, with the new implementation this accuracy is achieved already at $E^{\text{aux}}_{\text{cut}}=500 \,\text{eV}$ leading to $E_\text{ad}^{(2)} = 294.8\pm0.2\,\text{meV}$. This is at the cost of about $3\,770$ cpu hours which is only a small fraction of about 6\% compared to the cost of the previous code.} %Note that fitting error of the internal cutoff extrapolation is larger than the error of the manual cutoff extrapolation, since the former is performed for absolut MP2 energies, whereas the latter benefits from error cancelation effects.

%The internal cutoff extrapolation described in Sec. \ref{ch:CutoffExtrapol} is illustrated by an adsorption energy calculation of $\text{CH}_4$ in a chabazite crystal: the adsorption energy was computed against the plane wave cutoff $E_{\text{cut}}$. The lattice parameters and the orientation of the $\text{CH}_4$ molecule in a chabazite cage were taken from  \cite{Goltl2012} and then reoptimized using the optB88-vdW functional \cite{Klimes2010}. The entire system consists of 42 atoms in a unit cell of about 810 $\text{\AA}^3$. A visualization can be found in Fig. \ref{fig:Chabazite}. The total number of bands was set to 15872 where the first 100 are occupied bands. For the HF orbitals, plane wave cutoffs (\texttt{ENCUT} flag in \textsc{vasp}) of $450$, $550$, $650$, $750$, and $850$ eV were used, and all virtual orbitals were included in the  subsequent calculations (not to be confused with the plane wave cutoff, $E_{\text{cut}}$ (\texttt{ENCUTGW} flag in \textsc{vasp}), for the two-electron integrals in the MP2 calculation). The previous MP2 code clearly shows the mentioned $E_{\text{cut}}^{-3/2}$ behavior (\ref{eq:CutDecay}) as can be seen in Fig. \ref{fig:CutoffExtrapo}.  It is  worth mentioning that the MP2 curve reaches the converged value not before a cutoff of 1000 eV, if a tolerance of 1\% is assumed. However, the new LTMP2 algorithm attains this accuracy already at a cutoff of about 300 eV due to the internal extrapolation.

\section{Conclusion and Outlook}

We have presented an algorithm to calculate the exact MP2 energy for periodic systems, scaling only with the fourth power of the system size, $\mathcal O(N^4)$. The lower scaling is a consequence of a Laplace transformed energy denominator of the traditional MP2 formulation \revcolI{and of the use of Fast Fourier transforms}. In doing so the summations over the virtual bands can be carried out first, leading to transformed states in dependence of a plane wave index and a $\tau$-point. The loop over these plane waves is an outer loop that can be distributed over the CPUs without communication, leading to a very high parallelization efficiency. We showed that the parallelization is extremely close to ideal as long as the number of plane waves, $N_{\bm G}$, of the auxiliary basis set can be divided by the number, $\mathcal N_{\bm G}$, of parallelized plane wave groups. 

%The improved scaling and the lower memory requirements allow to calculate the MP2 energy of a $4\times 4 \times 4$ LiH supercell with 128 atoms on a modest system with only 64 CPUs in less than 19 hours. Based on this we estimate that the computation time for a $6 \times 6 \times 6$ supercell containing 432 atoms would be less than 2 hours if 100 000 CPUs are used. 

The slow convergence of the MP2 energy with respect to the number of basis functions is dealt with an extrapolation to an infinite basis set using the exact asymptotic cutoff behavior. \revcolI{In a comparison with the previous MP2 code in \textsc{vasp}, we demonstrated that this internal extrapolation leads to faster converging MP2 energies, reducing the computational effort significantly.}

In future the presented approach could be adapted to more involved electronic correlation energy methods, like second-order screened exchange (SOSEX) \cite{Freeman1977,Gruneis2009} or particle-hole ladder diagrams, in order to obtain a similar low complexity. Hence, the presented method can be considered as a step towards systematically improved correlation energies.

%Also, for calculating interatomic forces the concept of eliminating all summations over virtual states could be implemented leading to a lower scaling algorithm. As was shown by Ramberger et al. \cite{Ramberger2016}, forces can generally be computed using the self-energy in a Green's function approach. 

%Also, calculating interatomic forces should be possible using the presented low-complexity concept. Instead of the MP2 energy the self-energy at MP2 level has to be calculated. Having this, the interatomic forces can readily be computed with a Green's function approach that was published recently \cite{Ramberger2016}.

Also, calculating interatomic forces should be possible using the presented low-complexity concept, as the self-energy at MP2 level can be obtained along the same lines presented here for the MP2 energy. With the respective self-energy at hand, the recently published Green's function approach allows to efficiently calculate interatomic forces for perturbative methods, in this case MP2 \cite{Ramberger2016}.

%Also, calculating interatomic forces should be possible using a new Green's function and self-energy based concept which was published recently \cite{Ramberger2016}. Since the self-energy at the MP2 level has a similar form than the MP2 energy, the presented low-complexity approach which eliminates the summations over virtual bands could also be applied to MP2 forces

%Also, for calculating the self-energy at the MP2 level the presented low-complexity algorithm could be used. As shown recently \cite{Ramberger2016} this would lead to a 

%Also, for calculating interatomic forces at the MP2 level the presented low-complexity algorithm could be used. As shown recently by Ramberger and co-workers \cite{Ramberger2016}, the computation of forces only requires the independent particle Green's function and the self-energy. The self-energy at the MP2 level, however, has the same form as the  MP2 energy, such that presented low-complexity approach could be readily adapted for calculating the self-energy at the MP2 level.

%by providing the MP2 approximation of the self-energy. Forces are then calculated using the independent particle Green's function and the self-energy, as was shown recently by Ramberger and co-workers \cite{Ramberger2016} for the case of the randome phase approximation of the self-energy.

\section{Acknowledgements}

Computations were performed on the Vienna Scientific Cluster, VSC3. We also thank Ji\v{r}í Klime\v{s} for providing the optimized chabazite structure.

%%%%%%%%%%%%%%%%%%%%%%%%%%%%%%%%%%%%%%%%%%%%%%%%%%%%%%%%%%%%%%%%%
\section{Appendix}
%%%%%%%%%%%%%%%%%%%%%%%%%%%%%%%%%%%%%%%%%%%%%%%%%%%%%%%%%%%%%%%%%

\appendix

\section{Two-electron integrals\\in the plane wave basis}\label{A:TEFOPW}
For writing out a two-electron integral,
\begin{multline}
\langle i\bm k_1, j\bm k_2 | a\bm k_3, b\bm k_4 \rangle\\
= \int \D^3 r \int \D^3 r' \; \frac{\varphi_{i\bm k1}^*(\bm r) \varphi_{j\bm k_2}^*(\bm r')\varphi_{a\bm k_3}(\bm r)\varphi_{b\bm k_4}(\bm r') }{|\bm r - \bm r'|}  \;, \label{eq:2eIposition}
\end{multline}
in the plane wave basis we expand the Coulomb kernel, $1/|\bm r - \bm r'|$, in Fourier space:
\begin{equation}
\frac{1}{|\bm r - \bm r'|} = \frac 1 \Omega \sum_{\bm G}\sum_{\bm k}^{\text{BZ}} \frac{4\pi}{(\bm G + \bm k)^2} \EuE^{\ImI(\bm G + \bm k)(\bm r - \bm r')}\;. \label{eq:CoulombFourier}
\end{equation}
Note that in the strict sense this is only an approximation. However, in the thermodynamic limit, $N\rightarrow\infty$, $\Omega/N = \Omega_0 = \text{const.}$, the sum $1/\Omega \sum_{\bm k}$ turns into an integral $\int \D^3 k/(2\pi)^3$ and (\ref{eq:CoulombFourier}) becomes exact. Furthermore we use the identity
\begin{equation}
\sum_{\bm R} \EuE^{\pm \ImI(\bm k-\bm k') \bm R} = N \delta_{T(\bm k),T(\bm k')}\;. \label{eq:delta}
\end{equation}
Here $\sum_{\bm R}$ is a sum over all $N$ lattice vectors $\bm R$ of the system and $\bm k,\bm k'$ are crystal wave-vectors (not necessarily restricted to the first BZ). The function $T(\bm k)$ is defined on page \pageref{def:T}. Inserting (\ref{eq:CoulombFourier}) into (\ref{eq:2eIposition}), we split the integration over the entire system $\Omega$ into integrations over unit cells $\Omega_0$ translated by all lattice vectors:
%\begin{widetext}
%\begin{align}
%&\langle i\bm k_1, j\bm k_2 | a\bm k_3, b\bm k_4 \rangle \nonumber\\
%&=\frac 1 \Omega \sum_{\bm R\bm R'} \int_{\Omega_0}\D\bm r \int_{\Omega_0}\D\bm r'\sum_{\bm G}\sum_{\bm k}^{\text{BZ}} \frac{4\pi}{(\bm G + \bm k)^2} \EuE^{\ImI(\bm G + \bm k)(\bm r + \bm R - \bm r' -\bm R' )} \EuE^{-\ImI(\bm k_1 - \bm k_3)\bm R}\EuE^{-\ImI(\bm k_2 - \bm k_4)\bm R'} \varphi_{i\bm k_1}^*(\bm r) \varphi_{j\bm k_2}^*(\bm r')  \varphi_{a\bm k_3}(\bm r)  \varphi_{b\bm k_4}(\bm r') \nonumber \\
%&= \frac {N^2}{\Omega} \int_{\Omega_0}\D\bm r \int_{\Omega_0}\D\bm r'\sum_{\bm G}\sum_{\bm k}^{\text{BZ}} \frac{4\pi}{(\bm G + \bm k)^2} \EuE^{\ImI(\bm G + \bm k)(\bm r- \bm r')} \delta_{\bm k, T(\bm k_1 - \bm k_3)} \delta_{\bm k, T(\bm k_4 - \bm k_2)}\varphi_{i\bm k_1}^*(\bm r)   \varphi_{j\bm k_2}^*(\bm r') \varphi_{a\bm k_3}(\bm r)  \varphi_{b\bm k_4}(\bm r') \nonumber \\
%&= \frac {1}{\Omega} \delta_{T(\bm k_1 - \bm k_3), T(\bm k_4 - \bm k_2)} \sum_{\bm G} \frac{4\pi}{[\bm G + T(\bm k_1-\bm k_3)]^2}\langle i\bm k_1 | \EuE^{\ImI[\bm G + T(\bm k_1 - \bm k_3)]\hat{\bm r}} | a \bm k_3 \rangle_{\Omega_0} \langle j\bm k_2 | \EuE^{-\ImI[\bm G + T(\bm k_4 - \bm k_2)]\hat{\bm r}} | b \bm k_4 \rangle_{\Omega_0} \;.
%\end{align}
%\end{widetext}
\begin{gather}
\langle i\bm k_1, j\bm k_2 | a\bm k_3, b\bm k_4 \rangle \nonumber\\
=\nonumber\\
\frac 1 \Omega \sum_{\bm R\bm R'} \int_{\Omega_0}\D\bm r \int_{\Omega_0}\D\bm r'\sum_{\bm G}\sum_{\bm k}^{\text{BZ}} \frac{4\pi}{(\bm G + \bm k)^2} \EuE^{\ImI(\bm G + \bm k)(\bm r + \bm R - \bm r' -\bm R' )} \nonumber\\
\times \EuE^{-\ImI(\bm k_1 - \bm k_3)\bm R}\EuE^{-\ImI(\bm k_2 - \bm k_4)\bm R'} \varphi_{i\bm k_1}^*(\bm r) \varphi_{j\bm k_2}^*(\bm r')  \varphi_{a\bm k_3}(\bm r)  \varphi_{b\bm k_4}(\bm r') \nonumber \\
=\nonumber\\
\frac {N^2}{\Omega} \int_{\Omega_0}\D\bm r \int_{\Omega_0}\D\bm r'\sum_{\bm G}\sum_{\bm k}^{\text{BZ}} \frac{4\pi}{(\bm G + \bm k)^2} \EuE^{\ImI(\bm G + \bm k)(\bm r- \bm r')} \nonumber\\
\times\delta_{\bm k, T(\bm k_1 - \bm k_3)} \delta_{\bm k, T(\bm k_4 - \bm k_2)}\varphi_{i\bm k_1}^*(\bm r)   \varphi_{j\bm k_2}^*(\bm r') \varphi_{a\bm k_3}(\bm r)  \varphi_{b\bm k_4}(\bm r') \nonumber \\
=\nonumber\\
\frac {1}{\Omega} \delta_{T(\bm k_1 - \bm k_3), T(\bm k_4 - \bm k_2)} \sum_{\bm G} \frac{4\pi}{[\bm G + T(\bm k_1-\bm k_3)]^2} \nonumber\\
\times \langle i\bm k_1 | \EuE^{\ImI[\bm G + T(\bm k_1 - \bm k_3)]\hat{\bm r}} | a \bm k_3 \rangle_{\Omega_0} \langle j\bm k_2 | \EuE^{-\ImI[\bm G + T(\bm k_4 - \bm k_2)]\hat{\bm r}} | b \bm k_4 \rangle_{\Omega_0} \;. \nonumber
\end{gather}
In the first step the translated orbitals $\varphi_{a\bm k}(\bm r + \bm R) = \EuE^{\ImI\bm k \bm R}\varphi_{a\bm k}(\bm r)$ reveal phase factors which reduce to Kronecker deltas when summing over all lattice vectors, see Eq. (\ref{eq:delta}). The space integrals over $\Omega$ are thus reduced to integrals only over the unit cell $\Omega_0$. The Kronecker deltas eliminate the $\bm k$ sum over the BZ such that the Coulomb kernel leaves a sum only over the reciprocal lattice vectors. In the last step definition (\ref{eq:OverlapDens}) was adopted.

\section{Pseudocode for the parallelized implementation}

\begin{figure}
%\begin{minipage}[t]{\dimexpr 0.5\textwidth - 0.5\columnsep}
\includegraphics[width=0.85\linewidth]{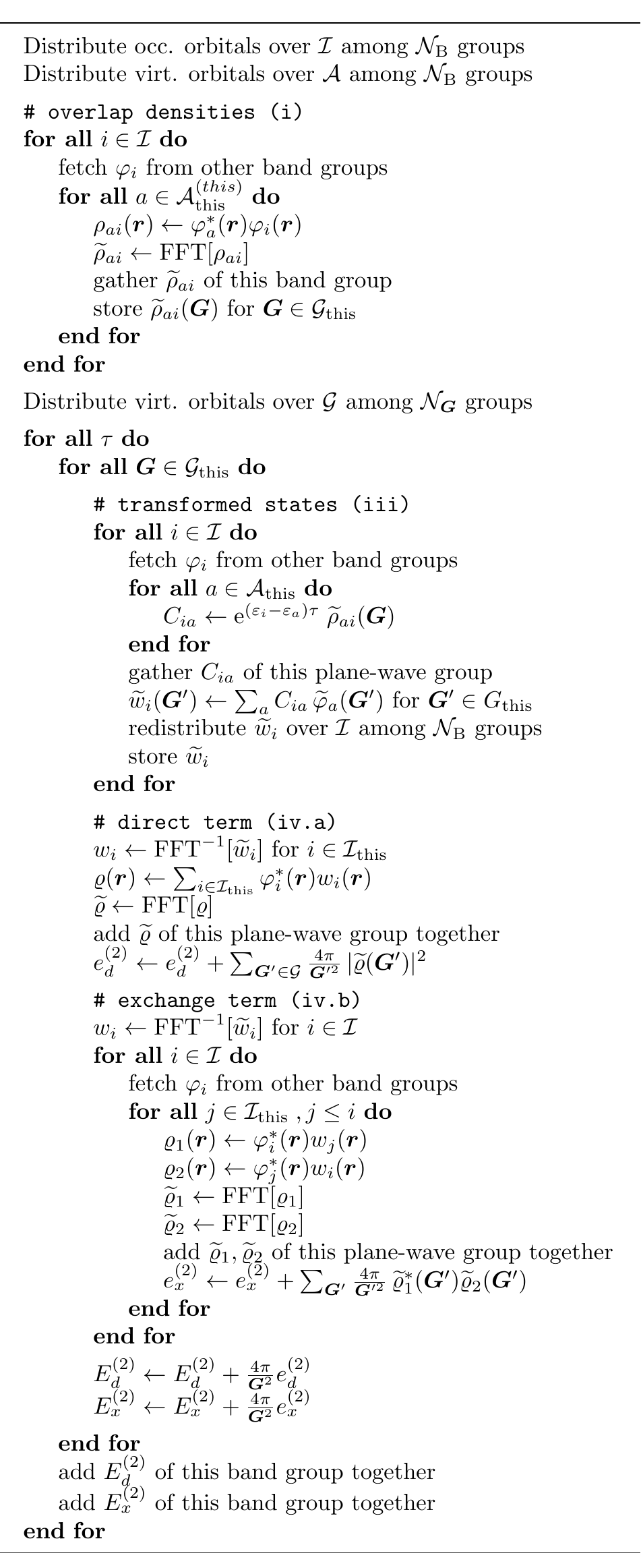}
\caption{Pseudocode of the parallel $\bm \Gamma$-only implementation. Hartree-Fock energies and orbitals for all occupied and virtual states are assumed.}
\label{fig:ParallelGamma}
%\end{minipage}\hfill
\end{figure}

\begin{figure}
%\begin{minipage}[t]{\dimexpr 0.5\textwidth - 0.5\columnsep}
%\vspace*{+2.0cm}
\includegraphics[width=0.7\linewidth]{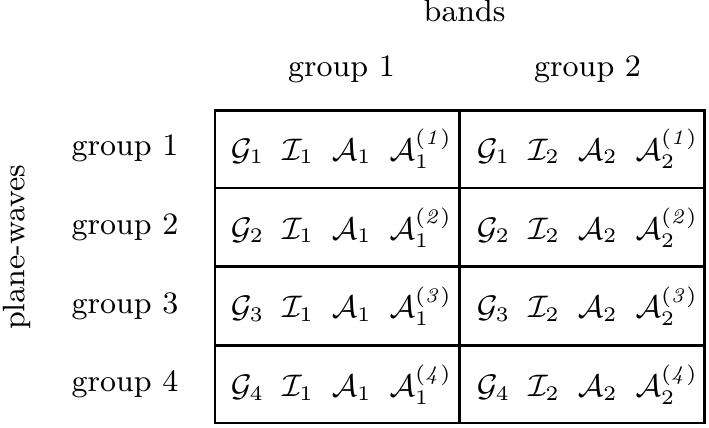}
%\vspace*{-0cm}
\caption{Illustration of the parallelization for the case of $\mathcal N = 8$ CPUs with $\mathcal N_{\text B} = 2$ band groups and $\mathcal N_{\bm G} = 4$ plane wave groups.}
\label{fig:ParallIllustration}
%\end{minipage}
\end{figure}

Pseudocode of the parallel $\bm \Gamma$-only implementation can be found in Fig. \ref{fig:ParallelGamma}. The number $\mathcal N$ of CPUs is divided into $\mathcal N_{\text B}$ groups parallelizing over bands and $\mathcal N_{\bm G}$ groups parallelizing over plane waves such that $\mathcal N = \mathcal N_{\text B} \mathcal N_{\bm G}$. The set $\mathcal G$ of all plane waves is evenly divided into $\mathcal N_{\bm G}$ disjoint subsets: $\mathcal G = \bigcup_n \mathcal G_n$, $n\in[1,\mathcal N_{\bm G}]$. Likewise the set of all occupied indices $\mathcal I$ and virtual indices $\mathcal A$ is evenly divided into $\mathcal N_{\text B}$ disjoint subsets: $\mathcal I = \bigcup_n \mathcal I_n$, $\mathcal A = \bigcup_n \mathcal A_n$, $n\in[1,\mathcal N_{\text B}]$. Moreover, each subset of virtual band indices $\mathcal A_n$ is further divided into $\mathcal N_{\bm G}$ disjoint subsets: $\mathcal A_n = \bigcup_m \mathcal A^{(m)}_n$, $m\in[1,\mathcal N_{\bm G}]$, since for the calculation of the overlap densities the plane wave parallelization is inapplicable. See Fig. \ref{fig:ParallIllustration} for an illustration.

%\newpage 

\bibliography{MP2}

\end{document}